\documentclass[aps,nofootinbib,amsmath,prd,onecolumn,notitlepage,showpacs,superscriptaddress,groupedaddress,11pt]{revtex4-1}

\usepackage[american]{babel}
\usepackage{amsfonts}
\usepackage{amsmath}
\usepackage{amssymb}
\usepackage{epsfig}
\usepackage{latexsym}
\usepackage{paralist}
\usepackage{fancyhdr}
\usepackage{graphicx}
\usepackage[vcentermath]{youngtab}
\usepackage{young}
\usepackage{ytableau}
\usepackage{etex}
\usepackage{braket}
\usepackage{float}
\usepackage{slashed}
\usepackage{xcolor}
\usepackage{soul}
\usepackage{dcolumn} 
\usepackage{bm}
\usepackage{nicefrac}
\usepackage{tipa}

\newcommand{\comment}[1]{ }

\def\bea{\begin{eqnarray}} 
\def\eea{\end{eqnarray}}
\def\be{\begin{equation}} 
\def\ee{\end{equation}} 
\def\ba{\begin{array}}
\def\ea{\end{array}}

\def\be{\begin{equation}}
\def\ee{\end{equation}}
\def\bea{\begin{eqnarray}}
\def\eea{\end{eqnarray}}

\usepackage{amsmath}

\begin{document}

\title{The origin of Weyl gauging in metric-affine theories}

\author{Dario Sauro}
\email{dario.sauro@phd.unipi.it}
\affiliation{
Universit\`a di Pisa and INFN - Sezione di Pisa, Largo Bruno Pontecorvo 3, 56127 Pisa, Italy}

\author{Omar Zanusso}
\email{omar.zanusso@unipi.it}
\affiliation{
Universit\`a di Pisa and INFN - Sezione di Pisa, Largo Bruno Pontecorvo 3, 56127 Pisa, Italy}

\begin{abstract}
%
In the first part, we discuss the interplay between local scale invariance and metric-affine degrees of freedom from few distinct points of view.
We argue, rather generally, that the gauging of Weyl symmetry is a natural byproduct
of requiring that scale invariance is a symmetry of a gravitational theory
that is based on a metric and on an independent affine structure degrees of freedom.
In the second part, we compute the N\"other identities associated with all the gauge symmetries,
including Weyl, Lorentz and diffeomorphisms invariances, for general actions
with matter degrees of freedom, exploiting a gauge covariant generalization of the Lie derivative. We find two equivalent ways to approach the problem, based on how we regard the spin-connection degrees of freedom, either as an independent object or as the sum of two Weyl invariant terms.
The latter approach, which rests upon the use of a Weyl-covariant connection with desirable properties,
denoted $\hat{\nabla}$, is particularly convenient
and constitutes one of our main results.
\end{abstract}

\maketitle

\renewcommand{\thefootnote}{\arabic{footnote}}
\setcounter{footnote}{0}

\section{Introduction}\label{sect:intro}

The idea that either rigid scale or local Weyl invariances could be fundamental
symmetries of nature has a very long history \cite{weyl1922space,Coleman:1970je,charap1974gauge,Smolin:1979uz}.
In fact, the idea has been resurrected multiple times and in various forms,
both for particle physics \cite{Meissner:2006zh,tHooft:2011aa,Ghilencea:2018thl}
and for the gravitational interactions \cite{Bekenstein:1980jw,Ghilencea:2015mza,Ghilencea:2018dqd}.
The most attractive feature of a scale or Weyl invariant spacetime
would be that the invariance could encompass the fundamental limitation
given by the Planck mass, that is, the dimensionful quantity that makes us think at
general relativity as an effective theory.
The assumption that spacetime is fundamentally scale invariant at high energies
could provide meaningful insights to a
theory of the traditional geometric degrees of freedom,
metric and connection, which works at arbitrarily high energies \cite{Mannheim:2011ds}.
Such theory could be
renormalizable, and even asymptotically free \cite{Fradkin:1985am} or safe \cite{Reuter:1996cp},
as long as scale invariance is broken in the low energy regimes \cite{Wetterich:2014gaa,Salvio:2017qkx}.
A description of geometry like the above would also cast gravity in a form that is closer
to the one of the other interactions of the standard model of particle physics,
which is scale invariant for the most part of its interactions.
Consequently, it would be a convenient feature
for a possible complete unification of forces \cite{tHooft:2010mvw}.

However, we need not forget that our understanding of general relativity,
and of the standard model for what matters, is, now more than ever, leaning
towards the fact that they are effective theories \cite{Baldazzi:2021kaf}, that work well on certain
intervals of scales, while eluding us outside of them.
Assuming that general relativity's metric and connection effective degrees of freedom
are also fundamental ones should be regarded as an endeavor, instead of
an obvious natural assumption.
The elusive part, that we discuss at length in this paper, is that if we try to enforce
in the same construction both the usual degrees of freedom of metric-affine gravity (MAG),
that is to say the aforementioned metric and connection as well as local scale invariance, we are lead naturally to a geometric theory in which
the metric-affine connection is complemented with a gauge potential for the Weyl symmetry. This could be seen as a theoretical prediction, or a price to pay,
depending on the reader's point of view.

In this paper, we discuss the naturalness of the inclusion of an Abelian potential
for the Weyl gauge symmetry in the context of general relativity and metric-affine gravity, and we give particular emphasis on torsion degrees of freedom and their conformal properties. Some of these aspects and their cosmological implications have already been investigated in the $f(T)$ literature, see e.g. \cite{Bahamonde:2015zma,Cai:2015emx,Krssak:2018ywd,Hohmann:2019nat,Gakis:2019rdd,Capozziello:2021pcg}, and the gauging of Weyl symmetry emerges quite naturally in the context of noncommutative gravity \cite{deCesare:2018cjr}. We give several arguments that motivate the form of the gauging of scale invariance \cite{Iorio:1996ad}, ranging from
the requirement that geodesics are not changed by a conformal transformation, to
the rethinking of the original Palatini's discussion
that a dynamical torsionless
connection must be Levi-Civita's on-shell \cite{Palatini:1919}. 
We find two equivalent ways of describing our metric-affine geometric setup,
one of which makes more transparent the Weyl covariance of the construction.
Bottom-up and top-down views on the naturalness of Weyl gauging are discussed in
Sects.~\ref{sect:bottom-up} and \ref{sect:top-down}, respectively,
while a direct connection with Palatini's approach to metric-affine
gravity is drawn in Sect.~\ref{sect:palatini}.
In the bottom-up approach, we show that the requirements of Weyl invariance and nonvanishing torsion force us to include a Weyl potential, enhancing the local gauge group from the Lorentz one to $SO(3,1) \times D(1)$, where $D(1)$
is the Abelian group of local Weyl rescalings. On the other hand, in the top-down appraoch, we start from a metric-affine viewpoint and analyze how the irreducible components of a general affine connection transform under Weyl rescalings. We eventually specialize to vanishing traceless nonmetricity, which allows us to focus on co-frame $e^a{}_\mu$, spin-connection $\omega^a{}_{b\mu}$ and Weyl potential $S_\mu$ as the natural gravitational field variables.

Armed with the necessary geometric toolkit, we also discuss the implications
that Weyl gauging, together with all other symmetries,
have on the coupling of matter fields
with gravity. We do so by obtaining the most general N\"other identities
associated with the coupling of matter fields on-shell
in terms of the currents that couple to the gauge potentials of our construction. These currents are energy-momentum tensor, dilation-vector and spin-current. 
We divide the derivation of N\"other identities in two parts, based on two equivalent but distinct approaches that come out naturally from the introductory discussion.
We refer to the approaches as the Cartan-Weyl, that features a conformally covariant torsion, and the Einstein-Weyl, that features instead a conformally invariant torsion, which are given in Sects.~\ref{sect:cartan-weyl} and \ref{sect:einstein-weyl}, respectively.
The main difference between the two viewpoints is that,
from the in Cartan-Weyl perspective the spin-connection is completely general, while from the Einstein-Weyl one the connection is split into two contributions $\omega^a{}_b = \hat{\omega}^a{}_b + \hat{\Omega}^a{}_b$. The splitting is chosen in such a way that $\hat{\omega}^a{}_b$ is a function only of co-frame and the Weyl gauge potential, $\hat{\omega}^a{}_b = \hat{\omega}^a{}_b (e^a, S)$,
transforming affinely under the local Lorentz group, while $\hat{\Omega}^a{}_b$ is a Lorentz tensor and is regarded as an independent field variable. We recover the traditional conservation laws of the energy-momentum tensor for vanishing dilation- and spin-currents.
Our analysis is complemented by an in-depth discussion of the relevant geometric
quantities and the use of a particular generalization of the Lie derivative,
introduced initially in Sect.~\ref{sect:improved},
which allows us to modify the generators of the diffeomorphis group
in a way that makes them covariant under all gauge symmetries,
including both local Weyl and Lorentz invariance.

The appendices contain further discussions of some geometrical aspects
that would have overburdenend the main text. Appendix~\ref{sect::AppendixCommutators}
includes relevant formulas for the commutators of the covariant derivatives
of the main text, the nontrivial contractions of the curvature tensors,
as well as the Bianchi identities associated to the
curvatures. Appendix~\ref{section-IntByParts} clarifies some aspects of the covariant integration by parts of connections in presence of Weyl gauging and including torsional degrees of freedom. Appendix~\ref{algebra} explores further the algebra associated to the covariant
Lie derivative that is used extensively in the main text.

\section{Weyl transformations vs independent connections: \\
a bottom-up approach}\label{sect:bottom-up}

In this section we present a bottom-up approach to Weyl gauging.
The presentation is going to be introductive and motivates the notion
of Weyl gauging as natural in the context of a formalism that accounts
simultaneously for an independent connection
(including, e.g., torsional degrees of freedom)
and for conformal Weyl rescalings of the metric.
This section also introduces much of the notation
that is necessary for the rest of the paper.

\subsection{Holonomic vs anholonomic degrees of freedom}\label{sect:mag}

To set the stage and part of the notation for the dicussion,
consider the two equivalent approaches towards general relativity
and metric-affine gravity:
the \emph{holonomic} approach
in which one works with a symmetric metric tensor $g_{\mu\nu}$
and a holonomic connection $\Gamma^\lambda{}_{\nu\mu}$,
and the \emph{anholonomic} approach (see e.g.\ \cite{1980ASIB...58..489C,Scholz:2018iuc,Kibble:1961ba,Gronwald:1995em}),
in which one works with a co-frame $e^a=e^a{}_\mu dx^\mu$
and a spin-connection $\omega^a{}_{b}=\omega^a{}_{b\mu} dx^\mu$.
The components of metric and co-frame
are related by the requirement that
$g_{\mu\nu}= \eta_{ab}e^a{}_\mu e^b{}_\nu$,
where $\eta_{ab}$ is the Minkowski metric.
We can switch from one approach to the other
by means of the tetrad postulate \cite{Yepez:2011bw},
which is the requirement that the full covariant derivative
of the co-frame vanishes
\begin{equation}\label{eq:connections}
 \nabla_\mu e^a{}_\nu =
 \partial_\mu e^a{}_\nu
 + \omega^{a}{}_{b\mu} e^b{}_\nu
 - \Gamma^\lambda{}_{\nu\mu} e^a{}_\lambda = 0\,,
\end{equation}
implying that the full connection is compatible with the co-frame.
We can use this equation to express either one of the two connections in terms of the vierbein, its inverse $E^\mu{}_a$, and the other connection.
For example, we have the relation
\begin{equation}\omega^{a}{}_{b\mu}
=E^\nu{}_a\Gamma^\lambda{}_{\nu\mu} e^a{}_\lambda-E^\nu{}_a \partial_\mu e^a{}_\nu\,.
\end{equation}
Notice that, in general, the holonomic-connection belongs to the algebra of
the group of general linear transformations,
$\Gamma_\mu\in gl(4)$, because
we are implicitly assuming that it is compatible with the metric,
$\nabla_\mu g_{\mu\nu}=0$, but it is not necessarily symmetric.
However, the requirement of metricity severely restricts the generality of the affine connection in that it can contain torsion components.
Likewise, $\omega^{a}{}_{b\mu}$ is also general,
yet, by definition,
it must belong to the adjoint representation of the Lorentz algebra,
$\omega_\mu \in so(3,1)$.
We anticipate that in the next section we are going to temporarily depart from the condition of metric compatibility to accommodate the effect of Weyl transformations, but also ultimately restore it introducing an additional
gauge component.

The two formulations are completely equivalent and simplify considerably
in the case of pure gravity with a Lagrangian density that is proportional to the scalar curvature, i.e.\ with the Einstein-Hilbert Lagrangian.
In this case, it can be shown that the field equations of the spin-connection yield the torsion-free condition (see, for example, Ref.\ \cite{gasperini2013theory}), which implies that the holonomic connection is symmetric.
In a similar way, a symmetric holonomic connection
is metric compatible on-shell \cite{Palatini:1919}. 
As a consequence, both connections can be expressed in terms of the vierbein and its derivatives in the anholonomic case, or the metric and its derivatives in the holonomic one.
We denote $\mathring{\nabla}_\mu$
the covariant derivative with components
$\mathring{\Gamma}^\mu{}_{\rho\nu}=\frac{1}{2}g^{\mu\lambda}(\partial_\nu g_{\lambda\rho}+\partial_\rho g_{\lambda\nu}-\partial_\lambda g_{\nu\rho})$,
that is the unique symmetric Christoffel connection, and $\mathring{\omega}^{a}{}_{b\mu}$, that is
the corresponding spin-connection obtained by the tetrad postulate
as in Eq.~\eqref{eq:connections}, $\mathring{\omega}^{a}{}_{b\mu}
=E^\nu{}_a\mathring{\Gamma}^\lambda{}_{\nu\mu} e^a{}_\lambda-E^\nu{}_a \partial_\mu e^a{}_\nu$.

In presence of some type of matter degrees of freedom,
especially spinors, the torsion $2$-form, denoted ${\cal T}^a$,
which is defined as the covariant curl of the co-frame,
${\cal T}^a=de^a + \omega^a{}_{b}\wedge e^b$,
might not vanish. As a consequence, the holonomic connection
is not generally symmetric unless we force it to be.
The connections can be written as before \emph{modulo} the contortion tensor
\begin{equation}\label{eq:connections-torsional-split}
  \omega^{a}{}_{b\mu}
  = \mathring{\omega}^{a}{}_{b\mu} + \Omega^{a}{}_{b\mu}
  \,,
  \qquad 
  \Gamma^\mu{}_{\rho\nu}
  = \mathring{\Gamma}^\mu{}_{\rho\nu} + K^\mu{}_{\rho\nu}
  \,.
 \end{equation}
In the above expressions,
$\Omega^{a}{}_{b\mu}$ is the torsional part of the spin-connection,
which is a tensor in both Lorentz and coordinate indices,
and $K^\lambda{}_{\nu\mu}$ is the contortion tensor.
The two are easily related by the tetrad postulate,
\begin{equation}\label{eq:omega-K-rel}
 \Omega^{a}{}_{b\mu}
=E^\nu{}_b K^\lambda{}_{\nu\mu} e^a{}_\lambda\,.
\end{equation}
The contortion tensor is related
to the components $T^{\mu}{}_{\nu\rho}= E^\mu{}_a {\cal T}^a{}(\partial_\nu,\partial_\rho) = 2\Gamma^\mu{}_{[\rho\nu]}$
of the torsion through the relation
$K^\lambda{}_{\mu \nu} = \frac{1}{2} \bigl( T_\mu{}^\lambda{}_\nu + T_\nu{}^\lambda{}_\mu - T^\lambda{}_{\mu\nu} \bigr)$.

\subsection{Weyl transformations vs independent connections}\label{sect:weyl}

Weyl tansformations, often referred to as conformal transformations in the literature of General Relativity \cite{Wald:1984rg}, are defined as the Abelian group of transformations in which the metric is rescaled by a conformal factor, $g_{\mu\nu} \to g'_{\mu\nu}= {\rm e}^{2 \sigma } g_{\mu\nu}$,
where $\sigma=\sigma(x)$ is a local function over spacetime.
The transformations are easily and unambiguously extended to the tetrads
\begin{equation}\label{Weyl-g}
  g_{\mu\nu} \rightarrow {\rm{e}}^{2 \sigma} g_{\mu\nu}\,, \qquad
  e^a{}_\mu \rightarrow {\rm{e}}^\sigma e^a{}_\mu\,, \qquad
  E^\mu{}_a \rightarrow {\rm{e}}^{- \sigma} E^\mu{}_a \,.
\end{equation}
The Christoffel connection and the associated spin-connection transform, for they are expressed in terms of metric's and co-frame's components
\begin{eqnarray}
 \mathring{\Gamma}^\mu{}_{ \rho\nu}
 & \rightarrow &
 \mathring{\Gamma}^\mu{}_{ \rho\nu}
 + \delta^\mu{}_\nu \partial_\rho \sigma
 + \delta^\mu{}_\rho \partial_\nu \sigma
 - g_{\nu\rho} g^{\mu\lambda} \partial_\lambda \sigma
 \,,\\
 \mathring{\omega}^{a}{}_{b\mu}
 & \rightarrow &
 \mathring{\omega}^{a}{}_{b\mu}
 + E^\nu{}_b e^a{}_\mu \partial_\nu \sigma
 - E^{\nu a} e_{b\mu}\partial_\nu \sigma\,.
\end{eqnarray}
One seemingly innocuous fact is that the Weyl transformation of the Christoffel connection contains three contributions, while the one of the associated spin-connection contains two.
This happens because $\mathring{\omega}^{a}{}_{b\mu}$ is antisymmetric
in the Latin indices and so must be its transformation.
To make it more transparent, it is sufficient to raise one index using
the metric $\mathring{\omega}^{ab}{}_\mu=\mathring{\omega}^a{}_{c\mu}\eta^{cb}$, in which case
\begin{eqnarray}
 \mathring{\omega}^{ a b}{}_\mu
 & \rightarrow &
 \mathring{\omega}^{ a b}{}_\mu
 +2 E^{\nu[a} \delta^{b]}{}_c e^c{}_{\mu} \partial_\nu \sigma\,,
\end{eqnarray}
where square brackets denote antisymmetrization with a factor.
Of course, this is a fundamental property if we want the transformation of
$\mathring{\omega}^{ a b}{}_\mu$ to be an element of the local Lorentz algebra
as it should. The same does not hold true for
$\mathring{\Gamma}^\mu{}_{\rho\nu}$,
which has a symmetric term proportional to $\delta^\mu{}_\rho$ in its transformation.

Now we must face the problem of how to extend the conformal properties to
the holonomic and anholonomic independent connections \cite{Iosifidis:2018zwo}.
There are two ``natural'' routes that we could follow.
The first one is to enforce that the holonomic connection transforms
like the Christoffel one, e.g.\ $\delta {\Gamma}^\mu{}_{\rho\nu} =
\delta^\mu{}_\nu \partial_\rho \sigma
+ \delta^\mu{}_\rho \partial_\nu \sigma
- g_{\nu\rho} g^{\mu\lambda} \partial_\lambda \sigma$, from which
we deduce that the contortion tensor must not transform,
$\delta K^\mu{}_{\rho\nu}=0$
(one index must be up).
Likewise, the torsional part of the spin-connection
does not transform either, $\delta \Omega^a{}_{b\mu}=0$.
As a consequence, conformal symmetry cannot be a symmetry
of the autoparallel equation,
$ \ddot{x}^\mu +\Gamma^\mu{}_{\rho\nu}\dot{x}^\nu\dot{x}^\rho=0$.
Geodesics of the resulting geometry will consequently change
according to the action of
the Weyl group.\footnote{%
Notice that here we are using
the symbol $\delta$ to denote finite transformations, not infinitesimal ones.} Note that even the covariant derivative of tensor that is not charged under Weyl would not be Weyl invariant anymore.

The second route, which is the one that we are concentrating on
the most in this paper,
is to require that the full connection does not transform,
$\delta {\Gamma}^\mu{}_{\rho\nu} =0$,
which does not change the autoparallel equation.
One straightforward reason to follow this route
is that, in the metric-affine formalism,
$g_{\mu\nu}$ and ${\Gamma}^\mu{}_{\rho\nu}$
are two independent objects, so, a priori, there is no reason why the Weyl
transformations of the two fields should be related with each other \cite{Iosifidis:2018zwo}.
In fact, this is the most natural
choice if we aim at constructing a conformal
(more precisely Weyl invariant) theory
that generalizes General Relativity above a certain energy scale
(e.g.\ the Planck mass). Such type of transformations are classified
in \cite{Iosifidis:2018zwo},
where they are called \emph{conformal transformation}.
However, now comes the crucial point: we are going to argue
that the presence of torsion and the requirement of Weyl invariance
are incompatible unless we promote the Weyl symmetry
from a local symmetry to a full gauge one.

\subsection{Affine Weyl transformation of the torsion tensor}

We have already written the transformation properties of the Christoffel symbols $\mathring{\Gamma}^\mu{}_{\rho\nu}$;
since ${\Gamma}^\mu{}_{\rho\nu}$ is required to be invariant,
the contortion tensor must transform in the opposite way as $\mathring{\Gamma}^\mu{}_{\rho\nu}$
\begin{equation}\label{Weyl-K}
 K^\mu{}_{\rho\nu} \rightarrow K^\mu{}_{\rho\nu}
 -\delta^\mu{}_\nu \partial_\rho \sigma
 - \delta^\mu{}_\rho \partial_\nu \sigma
 + g_{\nu\rho} g^{\mu\lambda} \partial_\lambda \sigma
 \,.
\end{equation}
Using the relation between the contortion tensor
and the torsionful part of the spin-connection, $\Omega^{a}{}_{b\mu}
=E^\nu{}_b K^\lambda{}_{\nu\mu} e^a{}_\lambda$,
we deduce the Weyl transformation of $\Omega_{\mu}{}^{a}{}_b$
\begin{equation}\label{Weyl-Omega}
 \Omega^{a}{}_{b\mu} \overset{?}{\to} \Omega^{a}{}_{b\mu}
 - e^a{}_\mu E^\rho{}_b \partial_\rho \sigma
 + e_{b\mu } E^{\rho a} \partial_\rho \sigma
 - \delta^a{}_b \partial_\mu \sigma\,.
\end{equation}
The origin of the incompatibility should now be transparent:
the last term of the above equation does not belong to the adjoint representation of the algebra of the local Lorentz group.
Instead, $\delta^a{}_b$ belongs to a $1$-dimensional trace-part of
a symmetric rank-$2$ tensor, rather than an antisymmetric one,
and the algebra is extended by one generator to an Abelian dilatation group,
say $D_1$. This feature of Weyl transformations has already been noticed in \cite{deCesare:2016mml}.

If we assume that fundamental fields must transform according to irreducible representations of the Lorentz group, we should amend the gauge transformation to eliminate the trace part. 
In order to do so, the straighforward option
is to introduce a new Abelian gauge potential, which we refer to as
the Weyl connection $S_\mu$ (see for example \cite{charap1974gauge}),
that accounts for covariance under
local scale symmetry and belongs to this representation.
We take $S_\mu$ to transform as
\begin{equation}\label{S-Weyl-transformation}
 S_\mu \rightarrow S_\mu - \partial_\mu \sigma \,,
\end{equation}
and we \emph{modify} the relation \eqref{eq:omega-K-rel}
between the contortion tensor and the torsional part
of the spin connection by replacing ${K}^\mu{}_{\rho\nu}$
with a new tensor ${\Phi}^\mu{}_{\rho\nu}$ as
\begin{equation}\label{Ktilda-def}
 \Phi^\mu{}_{\rho\nu}
  = E^\mu_a \Bigl( \Omega^{a}{}_{b\nu}  e_\rho^b + S_\nu e_\rho^a \Bigr)\,.
\end{equation}
Since we have changed the relation, we stress that $\Phi^\mu{}_{\rho\nu}$
does not have the same properties
as the contortion ${K}^\mu{}_{\rho\nu}$ that appeared in Eq.~\eqref{eq:connections-torsional-split}, as it is not antisymmetric in its first two indices.
We shall elaborate further on the actual difference in a moment.
We see that, even if we assume that the tensor
transforms as the old one (to balance $\mathring{\Gamma}^\mu{}_{\rho\nu}$),
$\delta \Phi^\mu{}_{\rho\nu}=-\delta^\mu{}_\nu \partial_\rho \sigma
 - \delta^\mu{}_\rho \partial_\nu \sigma
 + g_{\nu\rho} g^{\mu\lambda} \partial_\lambda \sigma$, now the trace part of \eqref{Weyl-Omega} is reproduced by \eqref{S-Weyl-transformation}, while $\Omega^{a}{}_{b\mu}$
remains an antisymmetric element of the algebra under transformations
\begin{equation}\label{Weyl-Omega-corrected}
 \Omega^{a}{}_{b\mu} \to \Omega^{a}{}_{b\mu}
 - e^a{}_\mu E^\rho{}_b \partial_\rho \sigma
 + e_{b\mu } E^{\rho a} \partial_\rho \sigma\,,
\end{equation}
which replaces \eqref{Weyl-Omega}.

The question is: what have we given up as compared to the discussion of Sect.~\ref{sect:mag}? Notice first the holonomic relation of the new connection
including $\Phi^\mu{}_{\rho\nu}$,
which we denote
\begin{equation}\label{eq:check-connection}
 \breve{\nabla}_\mu v^\nu = \partial_\mu v^\nu + \breve{\Gamma}^\nu{}_{\rho\mu} v^\rho
 = \partial_\mu v^\nu + \mathring{\Gamma}^\nu{}_{\rho\mu} v^\rho
 + K^\nu{}_{\rho\mu} v^\rho + S_\mu v^\nu\,,
\end{equation}
or, in other words, $\breve{\Gamma}^\nu{}_{\rho\mu}=\mathring{\Gamma}^\nu{}_{\rho\mu}+\Phi^\nu{}_{\rho\mu}=\mathring{\Gamma}^\nu{}_{\rho\mu}
+K^\nu{}_{\rho\mu} + S_\mu \delta^\nu{}_\rho$,
with $K^\nu{}_{\rho\mu}= E^\nu{}_a \Omega^a{}_{b\mu} e^b{}_\rho$
as in \eqref{eq:omega-K-rel}.
The new contribution to the connection is a special case of \emph{distortion}
tensor, $\breve{L}^\nu{}_{\rho\mu}=S_\mu \delta^\nu{}_\rho$, which implies that we have added nonmetricity to the original geometrical construction.
The nonmetricity tensor of $\breve{\nabla}$ is defined
$\breve{Q}_{\mu\nu\rho}=-\breve{\nabla}_\mu g_{\nu\rho} = 2 S_\mu g_{\nu\rho}$.

\subsection{Take one: Weyl gauging to restore compatibility of the metric}\label{sect:take-one}

Given the special form of the nonmetricity of the connection
introduced in \eqref{eq:check-connection} there is a natural
way to extend it to a metric compatible connection.
Assume that any field, say ${\cal V}^A$, where $A$ is an arbitrary collection
of holonomic and anholonomic indices, carries an Abelian charge
equal to its conformal weight $w({\cal V}^A)=w_{\cal V}$ (in units of length),
which is generally nonzero (notice that the weight depends on the covariance
properties of the tensor ${\cal V}^A$).
If we take the gauge field of the new Abelian group to be $S_\mu$
itself, i.e.\ the Weyl potential, then we can extend $\breve{\nabla}$
in \eqref{eq:check-connection} as
\begin{equation}\label{eq:tilde-connection-first-def}
 \tilde{\nabla}_\mu {\cal V}^A = \breve{\nabla}_\mu {\cal V}^A
 + w_{{\cal V}} S_\mu {\cal V}^A \,.
\end{equation}
The resulting connection is covariant under Weyl transformations:
for a transformation of the tensors
as ${\cal V}^A\to {\rm e}^{w_{{\cal V}}\sigma}{\cal V}^A$, 
it is straightforward to see that $\tilde{\nabla}_\mu{\cal T}^V\to {\rm e}^{w_{{\cal V}}\sigma}\tilde{\nabla}_\mu{\cal V}^A$.
In practice, covariant fields carry labels of both the local
Lorentz group and $GL(4)$ in $A$, as well as the Weyl group
(the latter being their weight), and we have gauged the latter.

Naturally, the metric $g_{\mu\nu}$ carries the Weyl charge
equal to its weight too, $w(g_{\mu\nu}) = w_g = 2$.
Using the definitions \eqref{eq:tilde-connection-first-def}
and \eqref{eq:check-connection} of the connections
and the weight, it is straightforward to show that
$\tilde{\nabla}$
is compatible with $g_{\mu\nu}$
\begin{equation}
 \tilde{\nabla}_\mu g_{\mu\nu} = 0 \,.
\end{equation}
In the next section we discuss a more top-down approach
to the construction of the same connection, which is both
more general and unveils independently Weyl covariant contributions
that are hidden in the general relation \eqref{eq:tilde-connection-first-def}.
The motivation for such an approach is that, even though the torsion that is
used in this section is Weyl covariant, it is not isomorphic (meaning in one-to-one correspondence) to the contortion tensor, which has an affine Weyl transformation. This feature comes from the torsion-vector contribution stemming from the presence of the Weyl potential in \eqref{eq:check-connection}.
In fact, in the approach of this section the torsion
transforms as a Weyl gauge potential, which
has allowed for applications in which the torsion
vector itself (modulo a constant) plays the role of $S_\mu$, see \cite{Karananas:2015eha,Karananas:2021gco,Karananas:2021zkl}.
Later on, we are also going to elaborate further on the properties of
the Weyl covariant connection.

\section{Weyl transformations vs independent connections: \\
the Weyl covariant formulation}\label{sect:top-down}

In the following we introduce a different and more convenient way to deal with torsion in presence of Weyl gauging (as compared to the procedure given in Sect.~\ref{sect:bottom-up}).
The approach provided here should also be regarded as a top-down point of view
on the construction of a Weyl covariant theory with MAG degrees of freedom.
In order to give some further insight on the origin of the gauging of the Weyl group, we spend some time surveying some features of metric-affine theories of gravity, so we temporarily reset our notation, but, in due time, we
reconnect with Sect.~\ref{sect:bottom-up}.

\subsection{MAG interlude: general connections and the requirement of Weyl invariance}

In MAG theories, the affine connection is an independent field variable from the onset. For our purpose, it is a $1$-form $\Gamma_\mu$ with values in the algebra of the gauge group $GL(4)$. Whenever a metric tensor $g_{\mu\nu}$ is defined on the spacetime manifold, the affine connection can be split as
\begin{equation}
\Gamma^\lambda{}_{\nu\mu} = \mathring{\Gamma}^\lambda{}_{\nu\mu} + \Phi^\lambda{}_{\nu\mu}\,,
\end{equation}
and $\mathring{\Gamma}$ is the symmetric compatible Levi-Civita connection, which is symmetric in the lower indices and metric compatible. Traditionally, the tensor $\Phi$
is further decomposed as $\Phi^\lambda{}_{\nu\mu}=N^\lambda{}_{\nu\mu}+K^\lambda{}_{\nu\mu}$, where $N$ is the distortion tensor and $K$ is the contortion tensor, which account for nonmetricity and torsion contributions, respectively.
The tensor $K^\lambda{}_{\nu\mu}$ is antisymmetric in the first indices and written only in terms of the torsion $T^\lambda{}_{\mu\nu}=\Gamma^\lambda{}_{\nu\mu} - \Gamma^\lambda{}_{\mu\nu}$, while $N^\lambda{}_{\nu\mu}$, which is sometimes called \emph{con-metricity} \cite{Floerchinger:2021uyo}, is symmetric in the lower indices and written solely in terms of the nonmetricity $Q_{\lambda\mu\nu}= - \nabla_\lambda g_{\mu\nu}$. The explicit forms are
\begin{subequations}
 \begin{align}\label{contortion-def}
  K^\lambda{}_{\nu\mu} & = \, \frac{1}{2} \left( T_\mu{}^\lambda{}_\nu + T_\nu{}^\lambda{}_\mu - T^\lambda{}_{\nu\mu} \right) \,, \\
  \label{distortion-def}
  N^\lambda{}_{\nu\mu} & = \, \frac{1}{2} \left( Q_\mu{}^\lambda{}_\nu + Q_\nu{}^\lambda{}_\mu - Q^\lambda{}_{\mu\nu} \right) \,.
 \end{align}
\end{subequations}

In order to understand the role of Weyl transformations on the components
of $\Phi$, however, we need a slightly different decomposition to begin with. The tensor $\Phi$, which is a rank-$3$ tensor under coordinate transformations,
can be split into irreducible parts with respect to the first two indices.
We focus on the first pair of indices because they provide the information on the subgroups of $GL(4)$ (different choices of splitting can be found in the literature, and the relationship with one of them will be discussed below).
Therefore, we perform the decomposition
\begin{equation}
\Phi_{\lambda\nu\mu} = \Phi_{(\lambda\nu)\mu} + \Phi_{[\lambda\nu]\mu}\,,
\end{equation}
where we also notice that we have made use of the metric to lower the first index.
The symmetric term can be expressed as the sum of trace and traceless parts, thence we have
\begin{equation}
\Phi^\lambda{}_{\nu\mu} = \delta^\lambda{}_\nu V_\mu + D^\lambda{}_{\nu\mu} + M^\lambda{}_{\nu\mu}\,.
\end{equation} 
We call $D_{\lambda\nu\mu} = D_{(\lambda\nu)\mu}$ the \emph{traceless distortion tensor}, owing to the property $D^\lambda{}_{\lambda\mu}=0$; $M_{\lambda\nu\mu} = M_{[\lambda\nu]\mu}$ is the \emph{generalized contortion tensor}; finally $V_\mu$ is the \emph{vector-distortion}.
We have that $M^\lambda{}_{\nu\mu}$ reduces to the usual contortion tensor in the limit in which both $D^\lambda{}_{\nu\mu}$ and the $V_\mu$ vanish, which should explain its name \cite{Hehl:1976kt}. Let us further split the traceless distortion tensor as
\begin{equation} \label{eq:traceless-dist-def}
D^\lambda{}_{\nu\mu} = \frac{2}{9} \left( B_\nu \delta^\lambda{}_\mu + B^\lambda g_{\mu\nu} - \frac{1}{2} \delta^\lambda{}_\nu B_\mu \right) + C^\lambda{}_{\nu\mu}\,,
\end{equation}
where the co-vector $B_\nu$ is defined as $B_\nu \equiv D^\lambda{}_{\nu\lambda}$ and all the traces of $C^\lambda{}_{\nu\mu}$ vanish identically. 

We can ask ourselves what are the consequences of requiring that the whole affine connection is Weyl invariant, similarly to the discussion of Sect.~\ref{sect:bottom-up}. In order to properly answer this question, we first need to introduce a conjugation in the algebra of $GL(4)$ induced by the metric
$\Phi^\lambda{}_{\nu\mu} \leftrightarrow g_{\nu\rho} g^{\lambda\kappa} \Phi^\rho{}_{\kappa\mu} $. This comes in handy to perform (anti)symmetrization of the first two indices, while preserving the covariance properties of $\Phi$ as a $GL(4)$ algebra element.
As in the previous section, we start from the assumption that the whole affine connection be Weyl invariant.
Focusing on the Weyl variation of the linear combinations of $\Phi$ and its conjugate we get
\begin{align}\label{eq:weyl-transf-phi-phiconj}
 \delta^W_\sigma \left(\Phi^\lambda{}_{\nu\mu} \pm g_{\nu\rho} g^{\lambda\kappa} \Phi^\rho{}_{\kappa\mu} \right) & = \, \delta^W_\sigma \Phi^\lambda{}_{\nu\mu} \pm g_{\nu\rho} g^{\lambda\kappa} \delta^W_\sigma \Phi^\rho{}_{\kappa\mu}  \\\nonumber
 & = \, -\delta^\lambda{}_\nu \left( 1 \pm 1 \right) \partial_\mu \sigma - \delta^\lambda{}_\mu \left( 1 \mp 1 \right) \partial_\nu \sigma + g_{\mu\nu} \left( 1 \mp 1 \right) \partial^\lambda \sigma\,.
\end{align}
Therefore, if we consider the symmetric combination (upper signs), we see that
it is natural to attribute the Weyl transformation
only to the trace part of the symmetric combination, i.e.\ to the vector distortion $V_\mu$, because the transformation is proportional to $\delta^\lambda{}_\nu$.
This implies naturally that $\delta^W_\sigma V_\mu = -\partial_\mu \sigma$
and $\delta^W_\sigma D^\lambda{}_{\nu\mu}=0$.
Similarly, if we consider the antisymmetric combination, we can read
the Weyl transformation of the generalized contortion tensor,
$\delta^W_\sigma M^\lambda{}_{\nu\mu}= g_{\mu\nu} \partial^\lambda \sigma-\delta^\lambda{}_\mu \partial_\nu\sigma$.
These choices ensure that the structure of the decomposition is unaltered
by the action of the Weyl transformation.\footnote{%
Notice that the covariance of the indices is important, as raising or lowering
them would change the action of $\delta^W_{\sigma}$. This is also
the reason why in \eqref{eq:weyl-transf-phi-phiconj}
we adopted that specific structure for the conjugate of $\Phi$.
}
Consequently, we see indirectly that the traceless distortion,
if present, is Weyl invariant if we demand that Weyl invariance is a
symmetry of the full affine connection.
We return shortly to the consequences of these equations.

Now notice that the generalized contortion term automatically drops if we write the covariant derivative of the metric tensor, i.e., the nonmetricity
\begin{equation}\label{eq:general-nonmetricity}
 Q_{\mu\nu\rho} \equiv - \nabla_\mu g_{\nu\rho} = 2 g_{\nu\rho} V_\mu + 2 D_{\nu\rho\mu} \,.
\end{equation}
The first term on the right hand side can be reabsorbed by a suitable redefinition of the covariant derivative, provided that we require the existence of a new $1$-dimensional, noncompact, Abelian symmetry group, in which case $V_\mu$ plays the role of its gauge connection as done in Sect.~\ref{sect:take-one}. The Weyl transformation of $V_\mu$ deduced above complies with this notion.
Therefore, the previous equation could be seen as merely expressing the need to write the covariant derivative of the metric tensor in a gauge-invariant way.
In contrast, the same logic cannot be extended to the traceless distortion tensor, which always gives rise to ``true'' nonmetricity.

Proceeding further, we can inspect the two inequivalent traces of the nonmetricity tensor
\begin{subequations}
\begin{align}
 Q_\mu{}^\lambda{}_\lambda = & \, 8 V_\mu \,, \\
 Q_\lambda{}^\lambda{}_\mu = & \, 2 V_\mu + 2 B_\mu \,.
\end{align}
\end{subequations}
Therefore, we could restate the previous result as follows. The trace of the nonmetricity in the second and third indices can be interpreted as the Abelian gauge-potential of a noncompact scale symmetry group, and similarly brought to the left hand side of $\nabla g = -Q$, making the whole expression gauge-invariant.

Now we focus our attention on the generalized contortion tensor. We denote its unique nontrivial contraction as $\chi_\nu \equiv M^\lambda{}_{\nu\lambda}$. We also separate the totally antisymmetric contribution defining $\theta_\nu \equiv M^\lambda{}_{\rho\mu} \varepsilon_{\nu\lambda}{}^{\rho\mu}$, and
introduce $\kappa^\lambda{}_{\nu\mu}$, which is the trace-free and not totally
antisymmetric part of the contortion tensor. We write the full decomposition as
\begin{equation}
 M^\lambda{}_{\nu\mu} = \frac{1}{3} \left( \chi_\nu \delta^\lambda{}_\mu - \chi^\lambda g_{\nu\mu} \right) + \frac{1}{6} \varepsilon_{\nu\mu}{}^{\lambda\rho} \theta_\rho + \kappa^\lambda{}_{\nu\mu}\,.
\end{equation}
This decomposition becomes useful if we concentrate on the torsion $2$-form, which,
in holonomic indices, is defined by
\begin{equation}
 T^\lambda{}_{\mu\nu} \equiv \Gamma^\lambda{}_{\nu\mu} - \Gamma^\lambda{}_{\mu\nu} = \left( M^\lambda{}_{\nu\mu} - M^\lambda{}_{\mu\nu} \right) + \left( V_\nu \delta^\lambda{}_\mu - V_\mu \delta^\lambda{}_\nu \right) + \left( D^\lambda{}_{\nu\mu} - D^\lambda{}_{\mu\nu} \right)\,.
\end{equation}
Notice that all the tensors defined from the generalized contortion also enter in the equation of the torsion $2$-form. To have a better insight on this feature, let us write down the expressions for the three irreducible components of the torsion tensor
(vector, axial-vector and tensor)
\begin{subequations}
\begin{align}
 \tau_\mu \equiv & \, T^\lambda{}_{\mu\lambda} =  - \chi_\mu - 3 V_\mu - B_\mu \,, \\
 \Theta_\nu \equiv & \, \varepsilon_{\nu\lambda}{}^{\rho\mu} T^\lambda{}_{\rho\mu} = \, 2 \theta_\nu \,, \\
 t^\lambda{}_{\mu\nu} = & \, \kappa^\lambda{}_{\nu\mu} - \kappa^\lambda{}_{\mu\nu} + C^\lambda{}_{\nu\mu} - C^\lambda{}_{\mu\nu}\,.
\end{align}
\end{subequations}
Even though only the generalized contortion enters the expression of the axial torsion-vector, distortion generally appears in the other two expressions.

In order to explicitly make contact with the previous section, we are going to discard the traceless distortion $D_{\mu\nu\rho}$ of the decomposition \eqref{eq:traceless-dist-def} henceforth.
In this limit, we can easily identify $V_\mu$ with the connection $1$-form $S_\mu$
that is associated to local scale and Weyl invariances, as argued before and done in Sect.~\ref{sect:bottom-up}.
Given the gauge-potential for scale transformations, which transforms as $S_\mu \rightarrow S_\mu - \partial_\mu \sigma$, we notice that we can make the following ``redefinition'' of the contortion
\begin{equation}\label{K-redefinition}
K^\lambda{}_{\nu\mu} = \hat{K}^\lambda{}_{\nu\mu } + \delta^\lambda{}_\mu S_\nu - g_{\mu\nu} S^\lambda\,.
\end{equation}
This definition has the advantage that a Weyl invariant term is singled out, in fact, the tensor $\hat{K}^\lambda{}_{\nu\mu }$ can thus be understood as a 
Weyl invariant generalization of the contortion tensor.

Using the Weyl invariant contortion,
the tensor part of the affine connection can be rewritten
\begin{equation}\label{Phi-redefinition}
\Phi^\lambda{}_{\nu\mu}
 = \delta^\lambda{}_\nu S_\mu + K^\lambda{}_{\nu\mu}
 = \delta^\lambda{}_\nu S_\mu + \delta^\lambda{}_\mu S_\nu - g_{\mu\nu} S^\lambda + \hat{K}^\lambda{}_{\nu\mu }
 \equiv L^\lambda{}_{\nu\mu} + \hat{K}^\lambda{}_{\nu\mu} \,,
\end{equation}
which \emph{defines} the tensor $L^\lambda{}_{\nu\mu}=\delta^\lambda{}_\nu S_\mu + \delta^\lambda{}_\mu S_\nu - g_{\mu\nu} S^\lambda$, see \cite{Smolin:1979uz,Cheng:1988zx}
(recall that we have set $D_{\mu\nu\rho}=0$ of \eqref{eq:traceless-dist-def} after having identified $S_\mu$). The original distortion $N^\lambda{}_{\nu\mu}$ given in \eqref{distortion-def} obviously differs from $L^\lambda{}_{\nu\mu}$.
 The tensor $L^\lambda{}_{\nu\mu}$ is precisely the contribution
that appears in the traditional construction of the gauged Weyl-covariant derivative,
which takes into account the fact that dilatations do not commute with spacetime transformations \cite{Iorio:1996ad,Karananas:2015eha}, though here it emerges from the decomposition under local scale transformations of a more general MAG connection.
The first term on the right hand side of \eqref{Phi-redefinition} is symmetric in the lower indices, as such it does not give rise to any torsion, whereas the second one is the Weyl invariant contortion tensor and can be used to find the torsion tensor itself. However, it is important to stress at this stage that $L^\lambda{}_{\nu\mu}$ is \emph{not} a distortion tensor, since it is not symmetric in the first two indices. Nevertheless, it contains a distortion part, i.e.\ the one which is proportional to $\delta^\lambda{}_\nu$.

\subsection{Take two: Weyl invariant torsion}\label{sect:take-two}

The discussion of Sect.~\ref{sect:bottom-up} on the bottom-up approach to
motivate the Weyl gauging procedure culminated in Sect.~\ref{sect:take-one}
where we assumed that the potential of the Abelian symmetry be the Weyl potential itself, extending the covariant derivative in \eqref{eq:tilde-connection-first-def}. It is natural to make the same choice in the second approach just outlined.
The added bonus coming from this section is that we have separated the Weyl
invariant contortion $\hat{K}$.

The new covariant derivative is defined as
\begin{equation}\label{eq:tilde-connection-second-def}
\tilde{\nabla}_\mu {\cal V}^A = \hat{\nabla}_\mu {\cal V}^A + (\hat{K}_\mu)^A{}_B {\cal V}^B 
+ w_{{\cal V}} S_\mu {\cal V}^A \,,
\end{equation}
where $w_{{\cal V}}$ is the Weyl weight of the tensor ${\cal V}^A$, and, again, $A$ is a collective multi-index standing for both Lorentz, coordinate and other internal indices. The notation $(\hat{K}_\mu)^A{}_B {\cal V}^B $ stands symbolically for the presence of a contortion contribution to the full covariant derivative (the components $\hat{K}^\nu{}_{\rho\mu}$ act on each index, as usual, with a plus sign for every contravariant coordinate index and a minus sign for every covariant one).

The connection is covariant under Weyl transformations:
for a transformation of the tensors
as ${\cal V}^A\to {\rm e}^{w_{{\cal V}}\sigma}{\cal V}^A$, 
it is straightforward to see that $\tilde{\nabla}_\mu{\cal V}^A\to {\rm e}^{w_{{\cal V}}\sigma}\tilde{\nabla}_\mu{\cal V}^A$.
As a consequence, covariant fields carry labels of both the local
Lorentz group in $A$ as well as the Weyl group, the latter being their weights.
In fact, the covariant derivative \eqref{eq:tilde-connection-second-def}
coincides with \eqref{eq:tilde-connection-first-def}.

The connection $\tilde{\nabla}_\mu$ is metric compatible and satisfies the tetrad postulate.
Metric compatibility follows easily from the antisymmetry of the contortion tensor in its first two indices and the definition of $L^\rho{}_{\nu\mu}$
\begin{equation} 
\tilde{\nabla}_\mu g_{\alpha\beta} = \mathring{\nabla}_\mu g_{\alpha\beta} - L^\lambda{}_{\alpha\mu} g_{\lambda\beta} - L^\lambda{}_{\beta\mu} g_{\alpha\lambda}
+ w_g S_\mu g_{\alpha\beta} =0\,,
\end{equation}
where we also used the original compatibility of the Levi-Civita connection $\mathring{\nabla}$ and the fact that the metric has weight two, $w_g=2$.

\subsubsection{Holonomic splitting and the covariant derivative $\hat{\nabla}$}

To give an explicit example of the holonomic action of $\tilde{\nabla}$, we take a spacetime vector $v^\mu$ with weight $w_v$. The connection acts as
\begin{eqnarray}\label{eq:nablahat-tilde-relation}
\tilde{\nabla}_\mu v^\nu
&=& \partial_\mu v^\nu+\tilde{\Gamma}^\nu{}_{\sigma\mu} v^\sigma
+ w_v S_\mu v^\nu
= \partial_\mu v^\nu+\hat{\Gamma}^\nu{}_{\sigma\mu} v^\sigma
+\hat{K}^\nu{}_{\sigma\mu} v^\sigma + w_v S_\mu v^\nu
\,.
\end{eqnarray}
The convenience of the approach discussed in this section should be apparent from the previous equation. The full holonomic connection $\tilde{\Gamma}^\nu{}_{\sigma\mu}$ is split in the torsion-free $\hat{\Gamma}^\nu{}_{\sigma\mu}$ and torsionful $\hat{K}^\nu{}_{\sigma\mu}$ parts, each of them being Weyl invariant.
In fact, we can write
\begin{eqnarray}
\tilde{\nabla}_\mu v^\nu
&=& \hat{\nabla}_\mu v^\nu
+\hat{K}^\nu{}_{\sigma\mu} v^\sigma 
\,,
\end{eqnarray}
where $\hat{\nabla}$, which does not include the 
torsional part $\hat{K}^\nu{}_{\sigma\mu}$, is the connection acting as
\begin{align}\label{eq:nablahat-definition}
 \hat{\nabla}_\mu v^\nu = \partial_\mu v^\nu + \hat{\Gamma}^\nu{}_{\rho\mu} v^\rho + w_v S_\mu v^\nu \,,
\end{align}
with components
\begin{align}
 \hat{\Gamma}^\nu{}_{\sigma\mu} = & \, \mathring{\Gamma}^\nu{}_{\sigma\mu} + L^\nu{}_{\sigma\mu}\\\nonumber
 = & \, \mathring{\Gamma}^\nu{}_{\sigma\mu} + \left( \delta^\nu{}_\sigma S_\mu + \delta^\nu{}_\mu S_\sigma - g_{\mu\sigma} S^\nu \right)\\\nonumber
 = & \, \frac{1}{2} \, g^{\nu\rho} \left( \hat{D}_\mu g_{\rho\sigma} + \hat{D}_\sigma g_{\rho\mu} - \hat{D}_\rho g_{\mu\sigma} \right)\,.
\end{align}
The connection $\hat{\nabla}$ becomes Weyl covariant thanks to the contribution of
the Weyl potential and reproduces \cite{Iorio:1996ad,Karananas:2015eha}. Among other things, this connection is symmetric
and compatible, but it can also be integrated by parts maintaining covariance (provided that the weight of the scalar that is integrated is $-4$, as it should to balance the weight of $\sqrt{-g}$, see Appendix \ref{section-IntByParts}). It can be thought of as a torsionless version
of $\tilde{\nabla}$, because they differ by the Weyl invariant contortion $\hat{K}$. We believe that the new procedure to obtain $\hat{\nabla}$,
as well as the procedure to couple it to torsional degrees of freedom, are the most important
results of our work, as we are trying to motivate in the following.

The components of the torsion of the affine connection $\tilde{\nabla}$ are
\begin{equation}
\tilde{T}^\sigma{}_{\mu\nu} =2 \tilde{\Gamma}^\sigma{}_{[\nu\mu]}
= 2 \hat{K}^\sigma{}_{[\nu\mu]} \, ,
\end{equation}
while the contortion tensor is written in terms of the torsion as in Eq.~\eqref{contortion-def}, i.e.\
\begin{equation}
 \hat{K}^\rho{}_{\nu\mu} = \frac{1}{2} \left( \tilde{T}_\nu{}^\rho{}_\mu + \tilde{T}_\mu{}^\rho{}_\nu - \tilde{T}^\rho{}_{\nu\mu} \right)\,.
\end{equation}
We also note that, defining the torsion-vector as
\begin{equation}\label{torsion-trace}
 \tilde{ \tau}_\mu \equiv \tilde{T}^\nu{}_{\mu\nu} \, ,
\end{equation}
we have
$
 \hat{K}^\nu{}_{\mu\nu} = - \tilde{ \tau}_\mu
$.

\subsubsection{Anholonomic splitting}

Up to now we have considered the splitting into torsion-free and torsionful pieces of the holonomic affine connection. We repeat the same reasoning for the anholonomic case as well.

Given the gauge transformations of the co-frame, the torsion $2$-form can be defined as its covariant exterior-derivative (i.e.\ its covariant-curl)
\begin{equation}
\tilde{T}^a = \tilde{D} e^a \, ,
\end{equation}
where $\tilde{D}$ is the part of $\tilde{\nabla}$ in which \emph{only} the gauge-potentials $\omega^a{}_{b\mu}$ and $S_\mu$ are considered.
In general, we adopt the notation in which the capitalized letter $D$
includes only the gauge connections, as opposed to $\nabla$ which contains
the full connection.

On the anholonomic side, and in absence of Weyl gauging, the splitting of the spin-connection, induced by the presence of a metric tensor, is usually written as $\omega^a{}_b = \mathring{\omega}^a{}_b +\Omega^a{}_b$ (see Sect.~\ref{sect:bottom-up}). Here $\mathring{\omega}^a{}_b$ solves the torsion-free condition $de^a + \mathring{\omega}^a{}_b \wedge e^b=0$, i.e., $\mathring{\omega}^a{}_b$ can be expressed in terms of $e^a$.
The torsion-free condition is Weyl covariant, because the infinitesimal Weyl transformation of $\mathring{\omega}^a{}_b$ takes the form $\delta^W_\sigma \mathring{\omega}^a{}_b = (\partial^b \sigma) e^a - (\partial^a \sigma) e^b $. Therefore, if we insist with the above splitting,
the Weyl covariant torsion $2$-form of the full Weyl invariant connection reads
\begin{equation}\label{eq:tilde-torsion-full}
\tilde{T}^a =
d e^a + (\mathring{\omega}+\Omega)^a{}_b \wedge e^b
+ S \wedge e^a
=\Omega^a{}_b \wedge e^b + S \wedge e^a\,.
\end{equation}
Since the last term is \emph{not} Weyl covariant, $\Omega^a{}_b$ itself cannot be Weyl invariant to balance out. In absence of Weyl gauging, this gives rise to a noncovariant torsion $2$-form \cite{Fabbri:2011vk}. The affine transformation properties of $\Omega^a{}_{b\mu}$ can be understood better by realizing that it is isomorphic
(meaning in one-to-one correspondence) to the noncovariant contortion tensor of Sect.~\ref{sect:bottom-up}, i.e., $\Omega^a{}_{b\mu} = e^a{}_\lambda E^\nu{}_b K^\lambda{}_{\nu\mu}$.

Similarly to the case of the full holonomic connection, we can take into account a more elaborate splitting, which is the anholonomic counterpart of \eqref{K-redefinition}:
\begin{equation}\label{omega-redefinition}
\omega^a{}_b = \mathring{\omega}^a{}_b + e^a (E_b \cdot S) - e_b (E^a \cdot S) + \hat{\Omega}^a{}_b \equiv \hat{\omega}^a{}_b+  \hat{\Omega}^a{}_b\,.
\end{equation}
We are going to see in a moment that $\hat{\Omega}^a{}_b$ is related to $\hat{K}$, $\hat{\Omega}^a{}_{b\mu} = e^a{}_\lambda E^\nu{}_b \hat{K}^\lambda{}_{\nu\mu}$ in anholonomic form.
Thus, using $e_b \wedge e^b = 0$, the covariant torsion $2$-form takes the form
\begin{equation}\label{eq:tilde-torsion-full-alt}
\tilde{T}^a = d e^a + \mathring{\omega}^a{}_b \wedge e^b + e^a \wedge S + \hat{\Omega}^a{}_b \wedge e^b + S \wedge e^a = \hat{\Omega}^a{}_b \wedge e^b\,,
\end{equation}
where we used the defining equation of $\mathring{\omega}^a{}_b$
and the antisymmetry of the wedge.
We have that $\hat{\Omega}^a{}_b$ is Weyl invariant, likewise $\hat{K}^\mu{}_{\nu\rho}$, as expected. We find a very similar expression, as well as the interpolating cases in Ref.~\cite{Izaurieta:2020kuy}, though the authors did not gauge the Weyl group, thus their results can be obtained from ours using the pure gauge limit of the Weyl potential ($S_\mu$ is a total derivative).

\subsubsection{General comments on dependences and compatibility in view of the N\"other identities}

Consider Eqs.~\eqref{eq:tilde-torsion-full} and \eqref{eq:tilde-torsion-full-alt}.
In both cases, the splitting can be seen as a simple redefinition of the components of the connections themselves, in which we rearranged the dependence on their parts.
In the first case, the three ``natural'' field variables become $(e^a,\Omega^a{},S)$, while in the second case they are $(e^a,\hat{\Omega}^a{}_b,S)$.
The two choices are possible, and both natural, to an extent. However, we claim that the second one is somewhat preferable, since $\hat{\Omega}^a{}_b$ is Weyl invariant and it is in one-to-one correspondence with the covariant torsion $2$-form, thus it can be more easily physically and geometrically interpreted. Clearly, Eq.~\eqref{omega-redefinition} is in complete analogy with Eq.~\eqref{K-redefinition} given above; in fact, it is straightforward to prove that there is actually an isomorphism,
\begin{equation}\label{Weyl-inv-contortion}
\hat{K}^\lambda{}_{\nu\mu} = E^\lambda{}_b e^a{}_\nu \hat{\Omega}^a{}_{b\mu}\,,
\end{equation}
relating the two tensors.

After the splitting of the spin-connection, the full covariant derivative of the co-frame takes the form
\begin{equation} \label{e-compatibility-with-tilde}
\tilde{\nabla}_\mu e^a{}_\nu = \partial_\mu e^a{}_\nu - \mathring{\Gamma}^\lambda{}_{\nu\mu} e^a{}_\lambda +  \hat{\omega}^a{}_{b\mu} e^b{}_\nu 
- \hat{K}^\lambda{}_{\nu\mu} e^a{}_\lambda - L^\lambda{}_{\nu\mu} e^a{}_\lambda + \hat{\Omega}^a{}_{b\mu} e^b{}_\nu
+ w_e S_\mu e^a{}_\nu \,,
\end{equation}
which vanishes for $w_e=1$ by virtue of the other compatibility, $\mathring{\nabla}_\mu e^a{}_\nu=0$, by use of Eq.~\eqref{Weyl-inv-contortion}, and
by the definitions of $\hat{\omega}^a{}_b$ and $L^\lambda{}_{\nu\mu}$ (see Eqs.~\eqref{K-redefinition}, \eqref{Phi-redefinition} and \eqref{omega-redefinition}).

\section{Weyl invariance and Palatini's approach} \label{sect:palatini}

A natural question is whether the lengthy discussions of Sects.~\ref{sect:bottom-up} and \ref{sect:top-down} are relevant in a more traditional
context closer to standard general relativity.
We briefly revisit here an argument that appeared recently
in Ref.~\cite{Wheeler:2022ggm}, adapting it to our purpose.
Consider the simplest Einstein-Hilbert action with independent variables
$g_{\mu\nu}$ and $\Gamma^\mu{}_{\nu\rho}$
\begin{equation}
S[g,\Gamma] = \int \sqrt{-g} g^{\mu\nu} R^\rho{}_{\mu\rho\nu}\,,
\end{equation}
where $R^\mu{}_{\nu\rho\theta}$ is the curvature tensor of $\Gamma^\mu{}_{\nu\rho}$,
which we take to be symmetric as in the Palatini's approach.
The variation of $S[g,\Gamma]$ with respect to the metric reproduces
Einstein's equations straightforwardly, but with the independent symmetric
connection.

In a famous paper \cite{Palatini:1919}, Palatini has shown that the metricity condition on $\Gamma^\mu{}_{\nu\rho}$ has dynamical origin.
The equations of motion coming from the variation of the connection are
\begin{equation}\label{Palatini1}
\nabla_\mu \left( \sqrt{-g} g^{\nu\rho} \right) = 0\,.
\end{equation}
The action of $\nabla_\mu$ on $\sqrt{-g}$ can be deduced by the fact that any operator, say $\delta$, that satisfies the Leibniz rule acts on the determinant as
$
\delta \sqrt{-g} =  -\frac{1}{2} \sqrt{-g} g_{\rho\nu} \delta g^{\rho\nu}
$.
Taking $\delta = \nabla_\mu$, we can rewrite the field equations \eqref{Palatini1} as
\begin{equation}\label{Palatini2}
\nabla_\mu g^{\nu\rho} - \frac{1}{2} g^{\nu\rho} g_{\lambda\sigma} \nabla_\mu g^{\lambda\sigma} = 0
\end{equation}
Notice that we are not extending the connection to densities,
but we are simply considering the determinant for what it is.
Contracting the previous equation with $g_{\nu\rho}$ and specializing to a manifold of dimension $d\neq 2$, one finds $g_{\nu\rho} \nabla_\mu g^{\nu\rho} = 0$, which can be inserted back in \eqref{Palatini2} to obtain $\nabla_\mu g^{\nu\rho} = 0$.
In short, we have that the connection has to be compatible with the metric on-shell
\begin{equation}
\nabla_\mu g_{\nu\rho}  = 0\,.
\end{equation}

The traditional interpretation is that the unique solution to the above equation
is $\Gamma^\mu{}_{\nu\rho}= \mathring{\Gamma}^\mu{}_{\nu\rho}$, that is, the Levi-Civita connection, which makes the Palatini action equivalent to Einstein-Hilbert one on-shell.
However, this is entirely true only in absence of Abelian gauge symmetries
under which the metric has a nontrivial charge.
To show this, assume that the metric tensor has some charge $w_g$
under an Abelian gauge symmetry and that it transforms as
\begin{equation}
g_{\mu\nu} \rightarrow {\rm e}^{w_g \sigma} g_{\mu\nu}\,.
\end{equation}
The Abelian group can be either compact or noncompact,
according to the value of $w_g$.
The relation with Weyl invariance should be obvious from the form of the transformation for $w_g=2$, in which case the group would be noncompact.

Denoting with $S_\mu$, as in the previous sections, the Abelian gauge potential
which transforms as
$S_\mu \rightarrow S_\mu - \partial_\mu \sigma$,
we have that the gauge covariant (under the Abelian group)
derivative of the metric tensor is
\begin{equation}
\hat{D}_\mu g_{\nu\rho} = \partial_\mu g_{\nu\rho} + w_g S_\mu g_{\nu\rho}\,.
\end{equation}
Recall that we reserve the capitalized letter $D$ for purely gauge covariant derivatives, which, in this case, contain the new Abelian charge $w_g$.
In fact, the latter equation is \emph{not} coordinate covariant,
because it must be supplemented by the connection $\Gamma^{\mu}{}_{\nu\rho}$
as in
\begin{equation}
\hat{\nabla}_\mu g_{\nu\rho} = \partial_\mu g_{\nu\rho}
- \Gamma^\lambda{}_{\nu\mu} g_{\lambda\rho}
- \Gamma^\lambda{}_{\rho\mu} g_{\nu\lambda}
+ w_g S_\mu g_{\nu\rho}\,,
\end{equation}
which is now covariant under both coordinate and Abelian transformations.

It is convenient to split $\Gamma^\mu{}_{\nu\rho}$ in terms of the Levi-Civita
connection $\mathring{\Gamma}^\mu{}_{\nu\rho}$ and an additional symmetric tensor,
i.e.~$\Gamma^\mu{}_{\nu\rho}=\mathring{\Gamma}^\mu{}_{\nu\rho}+L^\mu{}_{\nu\rho}$, which, to some extent, plays the role of a distortion. 
The tensor $L^\mu{}_{\nu\rho}$ inherits the symmetry property
of the original Palatini connection.
Looking back at the compatibility condition, we find
\begin{equation}
\hat{\nabla}_\mu g_{\nu\rho} \equiv \mathring{\nabla}_\mu g_{\nu\rho} - L^\lambda_{\phantom{\lambda}\nu\mu} g_{\lambda\rho} - L^\lambda_{\phantom{\lambda}\rho\mu} g_{\nu\lambda} + w_g S_\mu g_{\nu\rho} = 0\,.
\end{equation}
By construction, the Levi-Civita connection is compatible, $\mathring{\nabla}_\mu g_{\nu\rho}=0$, so we have
$ L_{\rho\mu\nu} + L_{\nu\mu\rho} -w_g S_\mu g_{\nu\rho} =0$.
We can apply the standard manipulation of the fundamental theorem
of Riemannian geometry (cycling the three indices, summing two expressions and subtracting the third) to find
\begin{equation}
L_{\rho\mu\nu} = \frac{w_g}{2} \left( S_\mu g_{\nu\rho} + S_\nu g_{\mu\rho} - S_\rho g_{\mu\nu} \right)\,.
\end{equation}
The ``hat'' connection that we have just written down coincides with the one given in Sect.~\ref{sect:take-two} if we assign the weight $w_g=2$
that is required for compatibility.
The full affine connection can be written in a manifestly gauge-invariant way
\begin{equation}
\Gamma^\rho_{\phantom{\rho}\mu\nu} = \frac{1}{2} g^{\rho\lambda} \bigl( \hat{D}_\mu g_{\nu\lambda} + \hat{D}_\nu g_{\mu\lambda} - \hat{D}_\lambda g_{\mu\nu} \bigr)\,,
\end{equation}
which represents a departure from the traditional Palatini solution,
$\Gamma^\mu{}_{\nu\rho}= \mathring{\Gamma}^\mu{}_{\nu\rho}$.
We stress that the departure is possible
only in presence of an additional Abelian symmetry
besides that of general covariance.

\section{N\"other identities in the Cartan-Weyl formalism}\label{sect:cartan-weyl}

Now we couple a matter field to our geometrical construction, the objective being
to find the general N\"other identities.
Consider an arbitrary matter field $\phi$ that transforms according to some linear representation of the Lorentz group and with a general Weyl weight.
We take a general Lorentz, Weyl and diffeomorphism invariant action of the form
\begin{equation}
 S_{\rm m}= S_{\rm m} [\phi, e^a, \omega^a{}_{b}, S]\,.
\end{equation}
We refer to this form as the \emph{Cartan-Weyl form} because the coframe and the gauge connections are regarded as independent, and the torsional degrees of freedom are
inside $\omega^a{}_{b}$, which has affine transformation properties under the Lorentz group. In the next section we discuss equivalent
identities that come from rearranging the dependencies of $S_{\rm m}$
in a different way.

The general functional differential of $S_{\rm m}$ can be expressed as
\begin{equation}\label{Variation}
 \delta S_{\rm m} =
 \int \frac{\delta S_{\rm m}}{\delta \phi} \delta \phi
 + \int \underline{e} \, T^\mu{}_{a} \delta e^a{}_\mu
 + \int \underline{e} \, \Sigma^\mu{}_{ab} \delta \omega^{ab}{}_\mu
 + \int \underline{e} \, \Delta^\mu \delta S_\mu \, ,
\end{equation}
where we have introduced the scalar density
$\underline{e} \equiv \det(e^a{}_\mu)=\sqrt{-g}$. Most importantly, we
define
the \emph{energy-momentum} tensor $T^\mu{}_a$, the \emph{spin-current} $\Sigma^\mu{}_{ab}$,
and the \emph{dilation-current} $\Delta^\mu$
as variations of $S_{\rm m}$ with respect to the gravitational fields,
\begin{equation}\label{tensors-definitions}
 T^\mu{}_a
 = \frac{1}{\underline{e}} \frac{\delta S_{\rm m}}{\delta e^a{}_\mu}
 \,,\qquad
 \Sigma^\mu{}_{ab}
 = \frac{1}{\underline{e}} \frac{\delta S_{\rm m}}{\delta \omega^{ab}{}_\mu}
 \,,\qquad
 \Delta^\mu
 = \frac{1}{\underline{e}} \frac{\delta S_{\rm m}}{\delta S_\mu}\,.
\end{equation}
The sign-convention of our definition of the energy-momentum tensor adheres with that of Weinberg's textbook \cite{Weinberg:1972kfs}. In the MAG literature, when the connection is completely general, the functional derivative of the matter action is referred to as \emph{hypermomentum} \cite{Hehl:1976kt}, but, in our formalism, it splits into the sum of the spin-current and the dilation-current.
On-shell, we can insert a solution to the equations of motion,
$\frac{\delta S_{\rm{m}}}{\delta \phi}=0$, thus canceling the first term of \eqref{Variation}. Then, the differentials can be evaluated
along the aforementioned symmetry transformations,
yielding the respective N\"other identities among the tensors.

\subsection{Local Lorentz and Weyl symmetries}

An infinitesimal local Lorentz transformation has the form $\Lambda^a{}_b=\delta^a{}_b+\alpha^a{}_b$, for a local antisymmetric matrix, $\alpha^{ab}= - \alpha^{ba}$. The transformation involves the Latin indices and is inhomogeneous
for the spin-connection.
The local Lorentz variations of co-frame, spin-connection and Weyl potential are
\begin{equation}\label{lorentz-transformations}
\begin{split}
 & \delta_\alpha^L \, e^a{}_\mu = \alpha^a{}_{b} e^b{}_\mu
 \,,\\
 & \delta_\alpha^L \, \omega^a{}_{b\mu}
 = - \partial_\mu \alpha^a{}_{b} - [\omega_\mu, \alpha]^a{}_{b}
 = - D_\mu \alpha^a{}_{b}
 \,,\\
 & \delta_\alpha^L \, S_\mu = 0\,.
\end{split}
\end{equation}
As in the previous sections, we have used the symbol $D_\mu$ to stress that the corresponding covariant
derivative is purely gauge, because, in general,
the transformation of a connection, such as $\omega^a{}_{b\mu}$,
can be rewritten as a gauge covariant
derivative acting on the transformation parameter itself (up to a convention-dependent sign).
Since the matrix $\alpha^a{}_{b}$ carries only Lorentz indices, we have
$D_\mu \alpha^a{}_{b}=\tilde{\nabla}_\mu \alpha^a{}_{b}$ by properly assigning Weyl weight zero to $\alpha^a{}_{b}$.

Using \eqref{lorentz-transformations} in \eqref{Variation}, going on-shell and integrating by parts, we find
\begin{equation}
\begin{split}
 \delta^L_\alpha S_{\rm m}
 &
 = \int \underline{e} \, T^\mu{}_{a} \alpha^a{}_{b} e^b{}_\mu
 - \int \underline{e} \, \Sigma^\mu{}_{ab} \tilde{\nabla}_\mu \alpha^{ab}
 =
 \int \underline{e} \, \alpha^{ab} \left\{ T^\mu{}_{a} e_{b\mu}
 + \tilde{\nabla}_\mu \Sigma^\mu{}_{ab} + \Sigma^\mu{}_{ab} \tilde{\tau}_\mu \right\} \,.
\end{split}
\end{equation}
The integration by parts of a nonmetric connection
has to be treated with care, so we refer to
the discussion of appendix \ref{section-IntByParts} for more insights.
From the above equation we see that Lorentz symmetry implies a generalized conservation law, which states that the antisymmetric part of the energy-momentum tensor is the divergence of the spin-current,
modulo a torsion-vector contribution \cite{Kibble:1961ba}.
Denoting $T_{ab}=T^\mu{}_{b} \eta_{ac} e^c{}_{\mu}$,
we have
\begin{equation}\label{anti-sym-energy}
 T_{[ab]} = (\tilde{\nabla}_\mu + \tilde{\tau}_\mu) \Sigma^\mu{}_{ab}\,.
\end{equation}
(see \cite{Kibble:1961ba}, as well as Eq.~(44) in \cite{Penrose:1983mf}, where the author deals with the Einstein-Cartan-Sciama-Kibble theory). 
The presence of an antisymmetric part is not unexpected, in fact, if the tensor $T_{\mu\nu}$ were computed using the connection $\mathring{\omega}^a{}_b$
the result would be symmetric on-shell, but the connection ${\omega}^a{}_b$
in this formulation is independent and does not guarantee such property.

Now we turn to Weyl gauge invariance.
We take the infinitesimal version of the Weyl transformations, seen also
in the previous sections, that is, $g_{\mu\nu}\to g'_{\mu\nu}=(1+2\sigma)g_{\mu\nu}$ for an infinitesimal local function $\sigma=\sigma(x)$.
The transformations are
\begin{equation}\label{weyl-transformations}
\begin{split}
 & \delta_\sigma^W e^a{}_\mu = \sigma e^a{}_\mu \,,
 \qquad
 \delta_\sigma^W \omega^a{}_{b\mu} = 0 \,,
 \qquad
 \delta_\sigma^W S_\mu = - \partial_\mu \sigma \,,
\end{split}
\end{equation}
where the fact that the spin-connection $\omega^a{}_{b\mu}$ is Weyl invariant reflects our assumption that the gauge-group is $SO(3,1) \times D(1)$,
i.e., a direct product (the same gauge transformations are taken into account in \cite{Karananas:2015eha}).
Inserting \eqref{weyl-transformations} in \eqref{Variation} with $\phi$
on-shell and integrating by parts, we find
\begin{equation}
 \delta_\sigma^W S_{\rm m}
 = \int \underline{e} \sigma \left\{ T^\mu{}_{a} e^a{}_\mu + (\tilde{\nabla}_\mu + \tilde{\tau}_\mu) \Delta^\mu \right\}\,.
\end{equation}
Furthermore,
using the definition of $\tilde{\nabla}_\mu$, given in \eqref{eq:nablahat-tilde-relation},
in terms of $\hat{\nabla}_\mu$, given in \eqref{eq:nablahat-definition}, and $\hat{K}_\mu$, we see that the vector-torsion contribution cancels when switching to $\hat{\nabla}_\mu$.
Thus, the consequence of Weyl gauge invariance is that the trace of the energy-momentum tensor
is the negative of the divergence of the dilation current
\begin{equation}\label{trace-energy}
 T^\mu{}_{\mu} = - \hat{\nabla}_\mu \Delta^\mu \,,
\end{equation}
if expressed in the Weyl covariant ``hatted'' connection.

Some comments are in order. Whereas Eq.~\eqref{anti-sym-energy} is widely known in the literature and represents the first step of the Belinfante procedure for the improvement of the energy-momentum tensor \cite{belinfante1940current},
Eq.~\eqref{trace-energy} usually appears in a different form, though we found a very similar expression in \cite{Floerchinger:2021uyo}.
This happens because Weyl symmetry is not generally \emph{gauged}, even though it
is a local symmetry either way.
Enforcing the invariance \emph{without} gauging leads to $T^\mu{}_{\mu}=0$,
which is a signature of Weyl symmetry
and also of conformal symmetry, to some extent.
Instead, $T^\mu{}_\mu$ is the divergence of a vector, modulo the torsion-vector term.
In the flat space limit, this is the signature of a scale invariant
(i.e.~a rigid Weyl invariant) theory \cite{ORaifeartaigh:1996hvx}.
The vector $\Delta^\mu$
for the case of a scale invariant theory in flat space is generally referred to as \emph{virial current} \cite{Coleman:1970je,Nakayama:2013is}.

\subsection{Diffeomorphism invariance and improved transformations}\label{sect:improved}

We finally turn our attention to the consequences of the diffeomorphism invariance of the theory. To discuss the last symmetry we find convenient to introduce
a different, yet equivalent, type of transformations,
which we refer to as ``improved'',
because they let us write down formulas that are covariant with repect to all the gauge symmetries at every step.
The diffeomorphism Einstein's variations are parametrized by a contravariant vector field locally written as
$\xi=\xi^\mu \partial_\mu$ and all gauge potentials transform infinitesimally
according to their Lie derivatives
\begin{equation}\label{einstein-transformations}
\begin{split}
 & \delta^E_\xi e^a{}_\mu = \pounds_\xi e^a{}_\mu \,,
 \qquad
 \delta^E_\xi \omega^a{}_{b\mu} = \pounds_\xi \omega^a{}_{b\mu}\,,
 \qquad
 \delta^E_\xi S_\mu = \pounds_\xi S_\mu\,.
\end{split}
\end{equation}
The algebra of infinitesimal diffeomorphisms satisfies the commutation relation
\begin{equation}
\begin{split}
 \left[\delta^E_{\xi},\delta^E_{\zeta}\right] = \delta^E_{[\xi,\zeta]}\,,
\end{split}
\end{equation}
where $[\xi,\zeta]$ are the Lie brackets of the two generators
with components
$[\xi,\zeta]^\mu
=\xi^\nu\partial_\nu \zeta^\mu-\zeta^\nu\partial_\nu \xi^\mu
= \xi^\nu\tilde{\nabla}_\nu \zeta^\mu-\zeta^\nu\tilde{\nabla}_\nu \xi^\mu
+  \xi^\nu \zeta^\rho \tilde{T}^\mu{}_{\rho\nu}
$, assuming that the vectors $\xi^\mu$ and $\zeta^\mu$ have Weyl weight zero.
The algebraic structure is a direct consequence of the same property holding for the Lie derivative.

\subsubsection{Improved transformations}

The Lie derivative is insensitive to the internal gauge indices of a given field or potential. As a consequence, the diffeomorphism transformation of tensors
that are not gauge singlets is not covariant with respect to the Lorentz group.
Take for example a spin vector $v^a$ with weight $w_v$, we obviously
have that the action of diffeomorphisms does not ``see'' the Latin index and the Weyl weight, $\delta^E_\xi v^a= \pounds_\xi v^a=\xi^\mu \partial_\mu v^a$.
For further insight, we can rewrite the partial derivative as the covariant one
and subtract the connection terms
\begin{equation}
\begin{split}
 \delta^E_\xi v^a = \xi^\mu \partial_\mu v^a
 &= \xi^\mu \tilde{\nabla}_\mu v^a
 - \xi^\mu \omega^a{}_{b\mu} v^b - w_v \xi^\mu S_\mu v^a \\
 &= \xi^\mu \tilde{\nabla}_\mu v^a
 - (\xi \cdot \omega)^a{}_b v^b - w_v (\xi \cdot S) v^a
 \,,
\end{split}
\end{equation}
where we introduced the shorthands $(\xi \cdot \omega)^a{}_b= \xi^\mu \omega_\mu{}^a{}_b$ and $\xi \cdot S= \xi^\mu S_\mu$.
It is an innocuous observation that the noncovariant terms can be written
as Lorentz and a Weyl transformations
\begin{equation}
\begin{split}
 \delta^E_\xi v^a = \xi^\mu \partial_\mu v^a
 &= \xi^\mu \tilde{\nabla}_\mu v^a
 - \delta^L_{\xi \cdot \omega} v^a - \delta^W_{\xi \cdot S} v^a
 \,,
\end{split}
\end{equation}
but the transformations are unusual in that their parameters contain
the connections of the respective gauge groups. The combination
$\delta^E_\xi v^a + \delta^L_{\xi \cdot \omega} v^a + \delta^W_{\xi \cdot S} = \xi^\mu \tilde{\nabla}_\mu v^a$ is covariant under all symmetry groups.
We are led to the definition of the following ``improved'' Einstein variations, which we distinguish over the other by a tilde
\begin{equation}\label{improved-Einstein-var}
 \tilde{\delta}^{E}_\xi \equiv \delta^E_\xi + \delta^L_{\xi \cdot \omega} + \delta^W_{\xi \cdot S} \,.
\end{equation}
which differ from the standard one by the terms that are necessary to make
the result covariant under all gauge groups.\footnote{%
If additional internal gauge groups with gauge potentials $A^i$ are present ($i$ runs over all the simple factors), then
Eq.~\eqref{improved-Einstein-var} can be naturally generalized to include them as $\tilde{\delta}^{E}_\xi \equiv \delta^E_\xi + \delta^L_{\xi \cdot \omega} + \delta^W_{\xi \cdot S} + \sum_i \delta^{G^i}_{\xi \cdot A^i}$.
}
Since the usual Einstein variations are the ordinary Lie derivatives, we are going to refer to the improved Einstein variations as the \emph{covariant Lie derivaties},
$\widetilde{\pounds}_\xi \equiv \tilde{\delta}^{E}_\xi$. Our definition is consistent with other applications in the metric-affine literature,
in which the covariant Lie derivative
is used to extend the Lie derivative to fields of arbitrary spin
by incorporating the local Lorentz factor \cite{Gronwald:1997jd,Obukhov:2006ge}.

One important remark is in order: even though the covariant Lie derivative is defined in a such a way that it yields (gauge) covariant quantities, it is different in nature from an ordinary covariant derivative and, in fact, cannot replace it. One simple way to see this is by noticing that it is not a directional derivative,
$\tilde{\pounds}_{\alpha \xi} \neq \alpha \tilde{\pounds}_{\xi}$ for $\alpha=\alpha(x)$ a scalar function over the spacetime. Directionality over the first argument is a
crucial for any covariant derivative. We elaborate more on the properties of $\tilde{\pounds}$ in appendix \ref{algebra}.

To begin with, we stress again that some of the gauge parameters of the improved transformation \eqref{improved-Einstein-var}
are not generic, but, instead, are the contraction
of the vector field $\xi^\mu$ with the gauge potentials.
For this reason, the transformations form an algebra, but not a Lie-algebra;
we will come back to this point after having studied the algebra of commutators.
With this choice, the improved Einstein variation for our protagonist fields are
\begin{equation}\label{impr-Einstein-def}
\begin{split}
 & \tilde{\delta}^{E}_\xi e^a{}_\mu
 = \pounds_\xi e^a{}_\mu
 + (\xi \cdot \omega)^a{}_{b} e^b{}_\mu + (\xi \cdot S) e^a{}_\mu \,,\\
 & \tilde{\delta}^{E}_\xi \omega^a{}_{b\mu}
 = \pounds_\xi \, \omega^a{}_{b\mu}
 - D_\mu (\xi \cdot \omega)^a{}_{b} \,,\\
 & \tilde{\delta}^{E}_\xi S_\mu =
 \pounds_\xi S_\mu - D_\mu(\xi \cdot S)\,,
\end{split}
\end{equation}
which, after some manipulation of the indices and using the tetrad postulate, can be rewritten as
\begin{equation}\label{impr-Einstein}
 \begin{split}
 & \tilde{\delta}^{E}_\xi e^a{}_\mu
 = \xi^\nu \tilde{T}^a{}_{\nu\mu} + e^a{}_\nu \tilde{\nabla}_\mu \xi^\nu
 \,,\\
 & \tilde{\delta}^{E}_\xi \omega^a{}_{b\mu} = \xi^\nu R^a{}_{b\nu\mu}
 \,,\\
 & \tilde{\delta}^{E}_\xi S_\mu = \xi^\nu W_{\nu\mu}
 \,.
 \end{split}
\end{equation}
In the above form the transformations highlight their geometrical origin
in terms of the curvature tensors (the torsion tensor can be thought
of as the curvature of the translations \cite{Scholz:2018iuc}).
It is straightforward to see that all the previous variations are both gauge- and coordinate-covariant.

Ultimately, if an action is invariant under the separate transformations
in \eqref{improved-Einstein-var}, then it is also invariant under \eqref{impr-Einstein}, so, assuming local Lorentz and Weyl invariance,
the invariance under the improved transformations is equivalent to the invariance
under traditional diffeomorphisms. We stress that the main use of covariant Lie derivatives is that of obtaining the associated N\"other identities in a full-covariant way.
Now we take a relatively long detour in discussing some geometrical properties
of the improved transformations,
the reader purely interested in their application
can skip directly to Sect.~\ref{sect:cartan-weyl-diff-noether}.

\subsubsection{Algebra of the improved transformations}

Now we turn our attention to the algebra of the new transformations; we are going to be succint with the computations and refer to Appendix \ref{algebra} for more details.
In order to understand the effect of the connection-dependent generators on the algebra of \eqref{improved-Einstein-var}, we first notice
that the Lorentz and Weyl subalgebras are ``twisted'' by the connections
\begin{equation}\label{impr-Einstein-twist}
 \left[\delta^L_{\xi\cdot\omega},\delta^L_{\zeta\cdot\omega}\right]
 = \delta^L_{[\xi,\zeta]\cdot \omega} + \delta^L_{{\cal R}(\xi,\zeta)}
 \,,
 \qquad
 \left[\delta^W_{\xi\cdot S},\delta^W_{\zeta\cdot S}\right]
 = \delta^W_{[\xi,\zeta]\cdot S}+\delta^W_{{\cal W}(\xi,\zeta)}
 \,,
\end{equation} 
where ${\cal R}(\xi,\zeta)^a{}_b=\xi^\mu\zeta^\nu R^a{}_{b\mu\nu}$
and ${\cal W}(\xi,\zeta)=\xi^\mu\zeta^\nu W_{\mu\nu}$. The right hand sides
of \eqref{impr-Einstein-twist} ``feel'' the presence of the curvature two-forms.
In particular, the twisted Weyl subalgebra is not even Abelian.

The commutator of two improved transformations can be computed with some work.
Using \eqref{impr-Einstein-twist} and \eqref{impr-Einstein-no-twist}, one can show that
\begin{equation}\label{cov-Lie-d-algebra}
 \bigl[ \widetilde{\pounds}_\xi, \widetilde{\pounds}_\zeta \bigr] = \widetilde{\pounds}_{[\xi,\zeta]} + \delta^L_{\mathcal{R}(\xi,\zeta)} + \delta^W_{\mathcal{W}(\xi,\zeta)}\,,
\end{equation}
where the first term of the right hand side is completely analogous to the one of the traditional transformation, while the additional two terms are implied by the twists
\eqref{impr-Einstein-twist}. Again, the commutator is sensitive to both Lorentz and Weyl curvatures.
In appendix \ref{algebra} we prove that the commutation rule \eqref{cov-Lie-d-algebra} holds for all fields, including gauge-connections, besides the tensors transforming covariantly under all gauge groups, so it is completely general.
In appendix \ref{algebra} we also prove that the Jacobi identities are satisfied for covariant Lie derivatives as well
\begin{equation}
 {\rm  Cycl }_{\xi,\zeta,\chi} \big[ \widetilde{\pounds}_\xi , \big[ \widetilde{\pounds}_\zeta, \widetilde{\pounds}_\chi \big] \big] = 0\,,
\end{equation}
which sums over the cycles of the three vectors. The algebra is thus closed,
as expected.

\subsubsection{Geometric interpretation \`a la Cartan}

The construction of the improved Einstein variations can be carried over in a manifestly diffeomorphism-invariant, geometric manner, using Cartan's formalism.
We start from the variation of the co-frame and make use of \emph{Cartan's magic formula} for Lie derivatives acting on $p$-forms, $\pounds_\xi = d \circ \iota_\xi + \iota_\xi \circ d$, where $\iota_\xi$ is the contraction of the $p$-form with $\xi$
from the left.\footnote{%
The action of $\iota_\xi$ can be defined iteratively as the $\mathcal{C}^\infty(M)$-linear map $\iota_\xi: \Omega^p(M) \rightarrow \Omega^{p-1}(M)$ that satisfies $\iota_\xi(dx^\mu)=\xi^\mu$ and $\iota_\xi (dx^\mu \wedge dx^\nu) = (\iota_\xi(dx^\mu)) dx^\nu - (\iota_\xi(dx^\nu)) dx^\mu $.
}
It is easy to see that the variation of $e^a{}_\mu$ in \eqref{impr-Einstein-def} can be written as
\begin{equation}
 \tilde{\delta}^E_\xi e^a
 = \left( \iota_\xi \circ d + d \circ \iota_\xi \right) e^a
 + (\iota_\xi \omega^a_{\phantom{a}b}) e^b
 + (\iota_\xi S ) e^a\,.
\end{equation}
The geometric interpretation can be unveiled using the following relations
\begin{equation}
\begin{split}
 (\iota_\xi S) e^a
  & = \iota_\xi ( S \wedge e^a)
  + S \, \iota_\xi e^a
  \,,\\
 (\iota_\xi \omega^a_{\phantom{a}b}) e^b
  & = \iota_\xi (\omega^a{}_{b} \wedge e^b)
  + \omega^a_{\phantom{a}b} \, \iota_\xi e^b\,,
\end{split}
\end{equation}
which highlight the analogy between the linear operators $\iota_\xi$ and $d$ acting on p-forms. Such relations can be exploited to obtain
\begin{equation}
 \tilde{\delta}^E_\xi e^a
 = \left( \iota_\xi \circ \tilde{D} + \tilde{D} \circ \iota_\xi  \right) e^a = \iota_\xi \tilde{\mathcal{T}}^a + \tilde{D} \xi^a \,,
\end{equation} 
where $\tilde{D}$ is the covariant exterior derivative on forms
that extends $d$.
The above equation is very interesting because it suggests us how to define, in general, a gauge-covariant Lie derivative of a given $p$-form in the fundamental representation of the gauge group. We simply replace the exterior-derivative with the covariant exterior-derivative. We find the same relation in \cite{Obukhov:2006ge}.

The same ``magic formula'' cannot hold for the Lie derivative of gauge-potentials, since it does not make sense to talk about their covariant exterior derivatives. Indeed, in the case of the spin-connection we find
\begin{equation}
\begin{split}
 \tilde{\delta}^E_\xi \omega^a{}_{b}
 & = \left( \iota_\xi \circ d + d \circ \iota_\xi \right) \omega^a{}_{b}
 - d \circ \iota_\xi (\omega^a{}_{b})
 - \omega^a{}_{c} \iota_\xi (\omega^c{}_{b})
 + \omega^c{}_{b} \iota_\xi (\omega^a{}_{c})
 = \iota_\xi \mathcal{R}^a{}_{b}\,,
\end{split}
\end{equation}
and similarly for the Weyl potential 
\begin{equation}
 \tilde{\delta}^E_\xi S = \iota_\xi \mathcal{W}\,.
\end{equation}

As a result of the change of basis \eqref{improved-Einstein-var} in the space of functional variations, the field strength $2$-forms have appeared on the right hand sides of \eqref{impr-Einstein}. This is a general feature, and we would see the same properties at work for other internal simple gauge factors as well, had we included them. Analogously, the covariant torsion $2$-form, which is the field strength of the co-frame, stands on the right-hand side of the variation of the co-frame itself \eqref{impr-Einstein}. As we have shown above, the same fact is easily understood in the geometric framework.

\subsubsection{Conservation laws for diffeomorphisms}\label{sect:cartan-weyl-diff-noether}

We have developed all the necessary machinery for taking into account the consequences of diffeomorphism invariance. Using \eqref{improved-Einstein-var} in \eqref{Variation} and integrating by parts, we find 
\begin{equation}
 \delta^{\tilde{E}}_\xi S
 = \int \underline{e} \, \xi^\nu \left\{
 - (\tilde{\nabla}_\mu + \tilde{\tau}_\mu) ( T^\mu{}_{a} e^a{}_\nu)
 + T^\mu{}_{a} \tilde{T}^a{}_{\nu \mu }
 + \Sigma^\mu{}_{ab} R^{ab}{}_{\nu\mu} + \Delta^\mu W_{\nu\mu} \right\}\,.
\end{equation}
The consequence is that the covariant divergence of the energy-momentum tensor in a theory invariant under both Lorentz and Weyl gauge symmetries reads
\begin{equation}\label{energy-cons}
 \tilde{\nabla}_\mu T^\mu{}_\nu
 = -\,\tilde{\tau}_\mu T^\mu{}_\nu + T^\mu{}_a \tilde{T}^a{}_{\nu \mu }
 + \Sigma^\mu{}_{ab} R^{ab}{}_{\nu\mu} +  \Delta^\mu W_{\nu\mu}\,.
\end{equation}
This is the generalization of the N\"other identity associated to diffeomorphism invariance for torsionful spacetimes \cite{Kibble:1961ba,Obukhov:2006ge} in presence of Weyl gauging. We found the analog of the previous equation in \cite{Neeman:1996zcr}, where it is derived in the metric-affine context, with the so-called metric energy-momentum playing the role of the dilation current and the non-metricity $1$-form that of the Weyl $2$-form.
Analogous metric-affine derivations of such N\"other identity can be found in \cite{Hehl:1994ue,Iosifidis:2020gth}.

Including another simple gauge factor in the internal gauge group simply adds a term $ j^\mu_I F^I_{\nu\mu}$ to the right hand side of \eqref{energy-cons}, where $ j^\mu_I \equiv \frac{1}{\underline{e}} \frac{\delta S_{\rm m}}{\delta A_\mu^I}$ is the gauge-current and $F^I_{\nu\mu}$ is the gauge field-strength with adjoint index $I$.
To gain more insight on the conservation law, we start by analyzing it in some limiting cases.

\subsubsection{Case $\Delta^\mu=\Sigma^{\mu\nu\rho}=0$}

As a first limit, let us consider the simplest situation in which
$\Delta^\mu=0$ and $\Sigma^{\mu\nu\rho}=0$, i.e.~the matter Lagrangian does not explicitly depend on the Weyl and Lorentz gauge potentials.
Under these assumptions, the energy-momentum tensor is automatically
symmetric and traceless thanks to the vanishing of the spin- and dilation-currents (see eqs.~\eqref{anti-sym-energy} and \eqref{trace-energy}). Since the energy-momentum tensor is symmetric, now Eq.~\eqref{energy-cons} reads
\begin{equation}
 \tilde{\nabla}_\mu T^{\mu\nu} = - \, \tilde{\tau}_\mu T^{\mu\nu} + \frac{1}{2}
 T^{\mu\rho} \left( \tilde{T}_{\rho}{}^\nu{}_{\mu}
 + \tilde{T}_{\mu}{}^\nu{}_{\rho} \right)\,.
\end{equation}
We further split the covariant-derivative on the left hand side singling-out the contortion contributions (symbolically $\tilde{\nabla}_\mu = \hat{\nabla}_\mu + \hat{K}_\mu$). Using $\hat{K}^\mu{}_{\nu\mu}= - \tilde{ \tau}_\nu$ and $T^{\mu\rho}=T^{\rho\mu}$, and the expression of the contortion tensor in terms of the covariant torsion, yields
\begin{align}
 \tilde{\nabla}_\mu T^{\mu\nu} & = \, \hat{\nabla}_\mu T^{\mu\nu}
 + \tilde{K}^\mu{}_{\lambda\mu} T^{\lambda\nu}
 + \tilde{K}^\nu{}_{\lambda\mu} T^{\mu\lambda}\\\nonumber
 & = \, \hat{\nabla}_\mu T^{\mu\nu} - \tilde{ \tau}_\mu T^{\mu\nu} + \frac{1}{2} T^{\mu\rho} \left( \tilde{T}_\rho{}^\nu{}_\mu + \tilde{T}_\mu{}^\nu{}_\rho \right)\,.
\end{align}
Therefore, in absence of the spin- and dilation-currents, the N\"other identity implied by the requirement of diffeomorphism-invariance is simply the conservation of the energy-momentum tensor from the viewpoint of the ``hatted'' symmetric affine connection
\begin{equation}
 \hat{\nabla}_\mu T^{\mu\nu} = 0
\end{equation}
Using again the symmetry of both the energy-momentum tensor and the affine connection, as well as $T^\mu{}_\mu=0$ and $w(T)=-6$, the previous relations becomes
\begin{align}\nonumber
 \hat{\nabla}_\mu T^{\mu\nu} = & \, \mathring{\nabla}_\mu T^{\mu\nu} + L^\mu{}_{\lambda\mu} T^{\lambda\nu} + L^\nu{}_{\lambda\mu} T^{\mu\lambda} - 6 S_\mu T^{\mu\nu} \\\nonumber
 = & \, \mathring{\nabla}_\mu T^{\mu\nu} +  4 S_\lambda T^{\lambda\nu} + 2 S_\mu T^{\mu\nu} - 6 S_\mu T^{\mu\nu} \\
 = & \,  \mathring{\nabla}_\mu T^{\mu\nu} \, ,
\end{align}
yielding automatically the usual conservation law of the energy-momentum tensor
\begin{equation}
 \mathring{\nabla}_\mu T^{\mu\nu} = 0\,.
\end{equation}

We remark the fundamental role of the trace-free nature of the energy-momentum tensor. If $T^{\mu\nu}$ is symmetric and trace-full and carries weight $w=-6$,\footnote{%
By construction, $T^\mu{}_\nu$ has weight $-4$, implying that the trace
is naturally integrated over with a weight $4$ density. Raising the lower index
as in $T^{\mu\nu}$ gives weight $-6$.
} it can be decomposed in the direct-sum $\underline{9} \oplus \underline{1}$, which reads
\begin{equation}
 T^{\mu\nu} = \Theta^{\mu\nu} + \frac{1}{4} g^{\mu\nu} \chi \, ,
\end{equation}
where $w(\chi)=-4$ and $\chi=T^\mu{}_\mu$. Thus, the ``hatted'' covariant derivative of such a tensor reads
\begin{equation}
 \hat{\nabla}_\mu T^{\mu\nu} = \mathring{\nabla}_\mu \Theta^{\mu\nu} + \frac{1}{4} g^{\mu\nu} \left( \partial_\mu \chi - 4 S_\mu \chi \right) \, ,
\end{equation}
In practice, the scalar component of the energy-momentum tensor feels the presence of the Weyl structure, whereas the $\underline{9}$-dimensional tensor component does not.

\subsubsection{Case $\Delta^\mu\neq 0$ and $\Sigma^{\mu\nu\rho}=0$}\label{subsect:with-dilation-no-spin}

The second limiting case that we want to study includes a nontrivial coupling
to the Weyl potential, but no interaction with the Lorentz one, that is,
$\Delta^\mu \neq 0$, $\Sigma^{\mu\nu\rho}=0$.
Now the energy-momentum tensor $T^{\mu\nu}$ is symmetric, but not necessarily traceless, though its trace is a total divergence.
Following similar steps, we find
\begin{equation}
 \hat{\nabla}_\mu T^{\mu\nu} = \Delta^\mu W^\nu{}_{\mu}\,.
\end{equation}
Therefore, in general, the usual local conservation of the energy and momentum $\hat{\nabla}_\mu T^{\mu\nu}=0$ is obtained only when the Weyl $2$-form strictly vanishes, i.e., when the Weyl potential is a pure gauge one. Splitting the contributions of the symmetric trace-free and trace-full parts of the energy-momentum tensor
\begin{equation}
T^{\mu\nu} = \Theta^{\mu\nu} + \frac{1}{4} g^{\mu\nu} \chi \, ,
\end{equation}
with $\chi= - \hat{\nabla}_\mu \Delta^\mu$. Further exploiting the results of Appendix \ref{sect::AppendixCommutators}, eqs.~\eqref{comm-rel-torsionfree} and \eqref{ContractionsAndRicci},
we can rewrite the conservation law as
\begin{equation}
 \hat{\nabla}_\mu \Theta^{\mu\nu}
 =
 \frac{1}{4} g^{\mu\nu} \left( \hat{\nabla}_\rho \hat{\nabla}_\mu \Delta^\rho -\widehat{Ric}_{\rho\mu} \Delta^\rho \right)\,.
\end{equation}
Thus, the conservation of the trace-free part of the energy-momentum tensor holds if and only if the right hand side vanishes identically, that is, for
\begin{equation}
 \hat{\nabla}_\mu \hat{\nabla}_\nu \Delta^\mu = \widehat{Ric}_{\mu\nu} \Delta^\mu
\end{equation}

\subsubsection{Case $\Delta^\mu=0$ and $\Sigma^{\mu\nu\rho}\neq 0$}\label{subsect:with-spin-no-dilation}

The last limiting case that we are focusing on is that of vanishing dilation-current and unknown spin-current. With these assumptions, Eq.~\eqref{energy-cons} reads
\begin{equation}
 \tilde{\nabla}_\mu T^\mu{}_\nu
 = -\,\tilde{\tau}_\mu T^\mu{}_\nu + T^\mu{}_a \tilde{T}^a{}_{\nu \mu }
 + \Sigma^\mu{}_{ab} R^{ab}{}_{\nu\mu}
\end{equation}
In presence of a nonvanishing spin-current, the energy-momentum tensor is asymmetric, the antisymmetric being given by Eq.~\eqref{anti-sym-energy}. On the other hand,    
in absence of dilation-current the symmetric part of the energy-momentum tensor is automatically trace-free. Thence, we can split $T^{\mu\nu}$ as
\begin{equation}
 T^{\mu\nu} = T^{(\mu\nu)} + T^{[\mu\nu]}
 = \Theta^{\mu\nu} + ( \tilde{\nabla} + \tilde{\tau} )_\rho \Sigma^{\rho\mu\nu} \, ,
\end{equation}
where $\Theta^{\mu\nu} \in \underline{9}$ and $T^{[\mu\nu]} \in \underline{6}$. Repeating the same expansion of the left hand side of the conservation equation considered in the special case of vanishing currents for $\Theta^{\mu\nu}$, we see that it still cancels out with the respective torsion-terms on the right hand side. Thus, the N\"other identity takes the form
\begin{equation}\label{cons-sigma-neq-0-1.2}
 \hat{\nabla}_\mu \Theta^{\mu\nu} + (\tilde{\nabla} + \tilde{\tau})_\mu  T^{[\mu\nu]} = \frac{1}{2} T^{[\mu\rho]} \left( \tilde{T}_\rho{}^\nu{}_\mu - \tilde{T}_\mu{}^\nu{}_\rho \right) + \Sigma^\mu{}_{\rho\lambda} \tilde{R}^{\rho\lambda}{}_{\nu\mu}\,.
\end{equation}
To gain further insight, we expand the covariant derivative of the antisymmetric part of the energy-momentum tensor
\begin{equation}
 \tilde{\nabla}_\mu T^{[\mu\nu]} = \hat{\nabla}_\mu T^{[\mu\nu]} - \tilde{\tau}_\mu T^{[\mu\nu]} - \frac{1}{2} \tilde{T}^\nu{}_{\rho\mu} T^{[\mu\rho]} \, ,
\end{equation}
which combines with both the vector-torsion contribution and the first term on the right hand side to give
\begin{equation}
  \hat{\nabla}_\mu \Theta^{\mu\nu} = - \hat{\nabla}_\mu  T^{[\mu\nu]} - \hat{K}_{\mu\rho}{}^\nu T^{[\mu\rho]}  + \Sigma^\mu{}_{\rho\lambda} \tilde{R}^{\rho\lambda}{}_{\nu\mu}\,.
\end{equation}
Exploiting again Eq.~\eqref{anti-sym-energy}, we finally obtain
\begin{align}
 \hat{\nabla}_\mu \Theta^{\mu\nu} = & \, - \hat{\nabla}_\lambda \hat{\nabla}_\mu \Sigma^{\lambda\mu\nu} - 2 W_{\lambda\mu} \Sigma^{\lambda\mu\nu} + \Sigma^{\lambda\mu\rho} \hat{R}_{\mu\rho}{}^\nu{}_\lambda \\\nonumber
 - & \, \hat{\nabla}_\mu \left( \hat{K}^\mu{}_{\rho\lambda} \Sigma^{\lambda\rho\nu} - \hat{K}^\nu{}_{\rho\lambda} \Sigma^{\lambda\rho\mu} + \hat{K}_{\lambda\rho}{}^\nu \Sigma^{\mu\lambda\rho} \right) + \Sigma^{\lambda\mu\rho} \hat{\nabla}^\nu \hat{K}_{\mu\rho\lambda}\,.
\end{align}
In this case the conservation equation looks more cumbersome; thus, in general, there is no straightforward limit in which the symmetric trace-free energy-momentum tensor $\Theta^{\mu\nu}$ is conserved.

\subsubsection{N\"other identities in geometrical language}\label{sect:geom-noether}

For the sake of completeness, we also give the N\"other identities stemming from Weyl, Lorentz and diffeomorphisms invariances in a geometric, coordinate-free notation. In this approach the gravitational field variables are given by the connection $1$-forms $\omega^a{}_b=\omega^a{}_{b\mu}dx^\mu$ and $S=S_\mu dx^\mu$, and by the co-frame $e^a=e^a{}_\mu dx^\mu$.
The associated curvature tensors are the Riemann $2$-form $\mathcal{R}^a{}_b= \frac{1}{2} R^a{}_{b\mu\nu} dx^\mu \wedge dx^\nu$, the Weyl (or homothetic curvature) $2$-form $\mathcal{W}=\frac{1}{2} W_{\mu\nu} dx^\mu \wedge dx^\nu$ and the torsion $2$-form $\tilde{\mathcal{T}}^a=\tilde{D} e^a$, where $\tilde{D}$ is the covariant exterior derivative.

In order to write the N\"other identities in the geometric language, we adopt the operation $\lfloor$, which contracts contravariant indices with covariant ones from the left. The general rule associated to such contraction can be obtained generalizing the following example:
let $\xi=\xi^\mu \partial_\mu$ be a vector field, then we can contract it with the Riemann $2$-form as
\begin{equation}
 \xi \lfloor \mathcal{R}^a{}_b \equiv \mathcal{R}^a{}_b(\xi,\,) = \xi^\mu R^a{}_{b\mu\nu} dx^\nu \, ,
\end{equation}
where in the second step only the first argument of the curvature $2$-form is evaluated on $\xi$, thus resulting in a $1$-form. We also use the dot symbol for contracting a covariant derivative with a contra-variant vector field, and denote the covariant divergence
\begin{equation}
 \hat{\nabla} \cdot \xi \equiv \hat{\nabla}_\mu \xi^\mu \, .
\end{equation}
Similarly we contract a $1$-form, for example $\tilde{ \tau}=\tilde{ \tau}_\mu dx^\mu$, with a vector field $\xi$
\begin{equation}
 \tilde{ \tau} \cdot \xi \equiv \tilde{ \tau}_\mu \xi^\mu \, .
\end{equation}

Having fixed the notation, it is time to come back to the physics. Given the matter action $S_m=S_m[\phi, e^a, \omega^a{}_b, S]$, where $\phi$ is the matter field as before, we have that the gravitational currents are the tensor-valued vector-fields defined in coordinate free notation as
 \begin{align}
  & T_a = \frac{1}{\underline{e}} \frac{\delta S_m}{\delta e^a} \,, \qquad
  \Sigma_a{}^b = \frac{1}{\underline{e}} \frac{\delta S_m}{\delta \omega^a{}_b} \, ,\qquad
  \Delta = \frac{1}{\underline{e}} \frac{\delta S_m}{\delta S} \,,
 \end{align}
in analogy with Eq.\ \eqref{tensors-definitions}.
Using these definitions, the N\"other identities associated in order to Weyl \eqref{trace-energy}, Lorentz \eqref{anti-sym-energy} and diffeomorphism invariance \eqref{energy-cons} can be written in the index-free form as
\begin{subequations}
 \begin{align}
  & \hat{\nabla} \cdot \Delta + T{}_a \lfloor \, e^a = 0 \, ,\\
  & (\hat{\nabla} + \tilde{ \tau}) \cdot \Sigma_{[ab]} + T{}_{[a} \lfloor \, e_{b]} = 0 \, ,\\
  &  (\tilde{\nabla} + \tilde{\tau} ) \cdot T{}_a + E_a \lfloor \, T{}_b \lfloor \, \tilde{\mathcal{T}}^b + E_a \lfloor \, \Sigma_{bc} \lfloor \, \mathcal{R}^{bc} + E_a \lfloor \, \Delta \lfloor \, \mathcal{W} = 0 \,.
 \end{align}
\end{subequations}

\section{N\"other identities in the Einstein-Weyl formalism} \label{sect:einstein-weyl}

One way to carry out the Belinfante improvement of the energy-momentum tensor \cite{belinfante1940current} is through the splitting of the spin-connection in two pieces: the first one, $\mathring{\omega}^a{}_b$, is the torsion-free spin-connection, while the other one, $\Omega^a{}_b$, is a tensor with Lorentz indices. Such a splitting is usually worked out requiring the tetrad postulate be fulfilled by two, separate, equalities.

Nevertheless, in our approach the independent object is $\hat{\Omega}^a{}_b$ -- see the discussion after Eq.~\eqref{eq:tilde-torsion-full-alt} -- whereas $\hat{\omega}^a{}_b$ depends on both the co-frame and the Weyl potential, yielding the Weyl invariant splitting discussed in Sect.~\ref{sect:top-down}
\begin{equation}
 \omega^a{}_{b\mu} = \hat{\omega}^a{}_{b\mu} + \hat{\Omega}^a{}_{b\mu}\,.
\end{equation}
The explicit expression of the ``hatted'' spin-connection $\hat{\omega}^a{}_b$ is
\begin{align}
 \hat{\omega}^a{}_{b\mu} = &  E^\nu{}_b \Bigl( \hat{\Gamma}^\lambda{}_{\nu\mu} e^a{}_\lambda - \partial_\mu e^a{}_\nu - S_\mu e^a{}_\nu   \Bigr)  \\\nonumber
 = &  E^\nu{}_b \Bigl( \mathring{\Gamma}^\lambda{}_{\nu\mu} e^a{}_\lambda - \partial_\mu e^a{}_\nu \Bigr) + \Bigl( e^a{}_\mu E^\nu{}_b - e_{b \mu } E^{\nu a} \Bigr) S_\nu\,.
\end{align}
The first contribution in the last line give rise to the Levi-Civita spin-connection $\mathring{\omega}^a{}_b$, whereas the remaining ones are the new contribution coming from the Weyl gauging.

For the purpose of obtaining the N\"other identities, we now want to take into account the implicit dependence of $\hat{\omega}^a{}_b$ on both the co-frame and the Weyl gauge-potential. Therefore, we need to compute the variation of $\hat{\omega}^a{}_b$, and we are going to start from the Levi-Civita piece. The variation of $\mathring{\omega}^a{}_b$ is given by
\begin{align}
\delta \mathring{\omega}^{ab}{}_\mu & = \frac{1}{2} E^{\nu b} \left( \mathring{\nabla}_\nu \delta e^a{}_\mu -  \mathring{\nabla}_\mu \delta e^a{}_\nu \right) + \frac{1}{2} E^{\nu a} \left(  \mathring{\nabla}_\mu \delta e^b{}_\nu - \mathring{\nabla}_\nu \delta e^b{}_\mu \right) \\\nonumber
& + \frac{1}{2} E^{\rho a} E^{\nu b} e_{c \mu} \left( \mathring{\nabla}_\nu \delta e^c{}_\rho - \mathring{\nabla}_\rho \delta e^c{}_\nu \right)\,.
\end{align}
The second term has the following variation
\begin{align}\nonumber
 \delta \left[ \left( e^a{}_\mu E^{\nu b} -  e^b{}_\mu E^{\nu a} \right) S_\nu \right] = & \, E^{\nu b} S_\nu \delta e^a{}_\mu - E^{\nu a} S_\nu \delta e^b{}_\mu + E^\nu{}_c E^{\rho a} e^b{}_\mu S_\nu \delta e^c{}_\rho - E^\nu{}_c E^{\rho b} e^a{}_\mu S_\nu \delta e^c{}_\rho \\
 + & \, \left( e^a{}_\mu E^{\nu b} -  e^b{}_\mu E^{\nu a} \right) \delta S_\nu\,.
\end{align}
After some manipulations and combining these results, we find
\begin{align}
 \delta \hat{\omega}^a{}_{b\mu } & =  \, \frac{1}{2} E^{\nu b} \left( \hat{\nabla}_\nu \delta e^a{}_\mu -  \hat{\nabla}_\mu \delta e^a{}_\nu \right) + \frac{1}{2} E^{\nu a} \left(  \hat{\nabla}_\mu \delta e^b{}_\nu - \hat{\nabla}_\nu \delta e^b{}_\mu \right) \\\nonumber
 & + \, \frac{1}{2} E^{\rho a} E^{\nu b} e_{c \mu} \left( \hat{\nabla}_\nu \delta e^c{}_\rho - \hat{\nabla}_\rho \delta e^c{}_\nu \right)\,.
\end{align}

As a consequence, the energy-momentum tensor $T^\mu{}_a$ -- see Eq.~\eqref{tensors-definitions} -- gets some correction terms stemming from the implicit dependence on $e^a$ in $\hat{\omega}^a{}_b$; likewise, the form of the dilation current $\Delta^\mu$ is modified. Using the definition of the spin-current $\Sigma^\mu{}_{ab}$ in \eqref{tensors-definitions}, we get the following contribution to the total variation of the matter action
\begin{align}
 \delta S_{\rm m} \supset \int \underline{e} \, \Sigma^\mu{}_{ab} \delta \hat{\omega}^{ab}{}_\mu = \int \underline{e} \, \left( \hat{\nabla}_\nu \Sigma^{\mu\nu\rho} + \hat{\nabla}_\nu \Sigma^{\rho\nu\mu} - \hat{\nabla}_\nu \Sigma^{\nu\mu\rho} \right) e_{a \rho} \delta e^a{}_\mu + 2 \int \underline{e} \, \Sigma^\nu{}_\nu{}^\mu \delta S_\mu\,.
\end{align}
The first two terms in round brackets modify the symmetric part of the energy-momentum term, whereas the third term contributes to the antisymmetric part.
Since we assume that our action functional depends on $e^a{}_\mu$, $\hat{\Omega}^{ab}{}_\mu$, $S_\mu$, besides some generic matter fields collectively denoted by $\phi$, a general variation with $\phi$ on-shell can be written as
\begin{align}\label{matter-variation-Einstein-Weyl}
 \delta S_{\rm m} = \int \underline{e} \,  \Theta^\mu{}_{a} \delta e^a{}_\mu + \int \underline{e} \,  \Sigma^\mu{}_{ab} \delta \hat{\Omega}^{ab}{}_\mu + \int \underline{e} \,  \mathfrak{D}^\mu \delta S_\mu\,,
\end{align}
where $\Theta^\mu{}_a$ is the full energy-momentum tensor, and $\mathfrak{D}^\mu$ is the full dilation-current. We remark that in general $\Theta^{\mu\nu}$ does not have definite symmetry properties, and it should not be confused with the symmetric trace-free tensor of the previous section. Since the splitting of the spin-connection into torsion-free and torsionful parts is linear, $\Sigma^\mu{}_{ab}$ remains unchanged
when compared to the previous section.

From the above calculations, we deduce that the improved energy-momentum tensor and dilation current are
\begin{subequations}\label{improved-tensors}
 \begin{align}
  \Theta^\mu{}_{a} & = T^\mu{}_{a} +  e_{a\rho} \left( \hat{\nabla}_\nu \Sigma^{\mu\nu\rho} + \hat{\nabla}_\nu \Sigma^{\rho\nu\mu} - \hat{\nabla}_\nu \Sigma^{\nu\mu\rho} \right)\\
  \mathfrak{D}^\mu & = \Delta^\mu + 2 \, \Sigma^\nu{}_\nu{}^\mu\, .
 \end{align}
\end{subequations}
We found an analogous improvement of the energy-momentum tensor in \cite{Hehl:1994ue,Iosifidis:2020gth,Floerchinger:2021uyo}, in the more general arena of metric-affine theories. In \cite{Hehl:1994ue} such improvement is derived in the geometric language of differential forms, whereas in \cite{Iosifidis:2020gth,Floerchinger:2021uyo} the derivation is coordinate-based.

We must stress that the energy-momentum tensor is best written in holonomic indices, defining $\Theta^\mu{}_\nu = \Theta^\mu{}_a e^a{}_\nu $. The reason is that we know
that the energy-momentum tensor must become an energy density, i.e.~it must give the energy per unit volume at a given spacetime point. As such, it has mass dimension $4$, implying that its Weyl weight is $-4$, which is precisely weight with holonomic indices is $\Theta^\mu{}_\nu$ (one index up and one down). The above form of the energy-momentum tensor is also consistent with the result usually given in classical field theory, i.e.
\begin{equation*}
 \Theta^\mu{}_\nu = \frac{\partial \mathcal{L}}{\partial (\partial_\mu \phi^i)} \partial_\nu \phi^i - \delta^\mu{}_\nu \mathcal{L}\,.
\end{equation*}
For completeness, we notice that the same logic would allow us to consider also $\Theta^a{}_b$, for $w(\Theta^a{}_b) = w(\Theta^\mu{}_\nu)$. We are not going to speculate further on this choice in the present paper.

\subsection{Lorentz and Weyl invariances}

Having derived the total variation of the matter action with matter fields on-shell, and having defined the improved currents associated with the three field variables $e^a{}_\mu$, $\hat{\Omega}^a{}_{b\mu}$ and $S_\mu$, we are ready to inspect the N\"other identities implied by Lorentz, Weyl and diffeomorphisms invariances. We start with Lorentz invariance as in the previous section, the only difference with respect to \eqref{lorentz-transformations} is that $\hat{\Omega}^a{}_b$ is a tensor under Lorentz indices, thus
\begin{equation}\label{lorentz-transformations-2}
\begin{split}
& \delta_\alpha^L \, e^a{}_\mu = \alpha^a{}_{b} e^b{}_\mu
\,,\\
& \delta_\alpha^L \, \hat{\Omega}^a{}_{b\mu}
=  \alpha^a{}_c \hat{\Omega}^c{}_{b\mu} - \alpha^c{}_b \hat{\Omega}^a{}_{c\mu}
=  [\alpha, \hat{\Omega}_\mu]^a{}_{b}
\,,\\
& \delta_\alpha^L \, S_\mu = 0\,.
\end{split}
\end{equation}
Using these variations in the first equation of \eqref{improved-tensors} yields
\begin{equation}
 \delta^L_\alpha S_{\rm m} = \int \underline{e} \alpha^{ab} \left\{ \Theta^\mu{}_a e_{b \mu} - \Sigma^\mu{}_{ac} \hat{\Omega}^c{}_{b\mu} + \Sigma^\mu{}_{bc} \hat{\Omega}^c{}_{a\mu} \right\}\,.
\end{equation}
Therefore, Lorentz invariance imposes the following constraint on the antisymmetric part of the improved energy-momentum tensor
\begin{equation}\label{anti-sym-theta1}
 \Theta_{[\mu\nu]} = \hat{K}^\lambda{}_{\mu\rho} \Sigma^\rho{}_{\nu\lambda} - \hat{K}^\lambda{}_{\nu\rho} \Sigma^\rho{}_{\mu\lambda} = \tilde{\nabla}_\rho \Sigma^\rho{}_{\mu\nu} - \hat{K}^\rho{}_{\lambda\rho} \Sigma^\lambda{}_{\mu\nu} - \hat{\nabla}_\rho \Sigma^\rho{}_{\mu\nu}\,.
\end{equation}
On the other hand, from Eq.~\eqref{improved-tensors}, we observe that
\begin{equation}\label{anti-sym-theta2}
 \Theta_{[\mu\nu]} = T_{[\mu\nu]} - \hat{\nabla}_\rho \Sigma^\rho{}_{\mu\nu} \,.
\end{equation}
Using $\hat{K}^\rho{}_{\lambda\rho}= - \tilde{\tau}_\lambda$, which relates the trace of the contortion to the vector torsion, Eq.~\eqref{anti-sym-theta1} becomes
\begin{equation}
 \Theta_{[\mu\nu]} = \left( \tilde{\nabla}_\rho + \tilde{\tau}_\rho \right) \Sigma^\rho{}_{\mu\nu} - \hat{\nabla}_\rho \Sigma^\rho{}_{\mu\nu}\, ,
\end{equation}
which is fully consistent with Eq.~\eqref{anti-sym-energy}. The consequence is that imposing Lorentz invariance in the Einstein-Weyl formalism or in Cartan's approach is completely equivalent, as we could have naively expected. We remark that, in both cases, the energy-momentum tensor is symmetric if the hypermomentum current is zero,
$\Sigma^{\rho\mu\nu}=0$.

Moving on to Weyl symmetry,
the variations of the field variables under Weyl gauge transformations are exactly the same as in the Cartan formalism -- see Eq.~\eqref{weyl-transformations} -- thanks to the fact that we have built $\hat{\Omega}^a{}_b$ in such a way that it is Weyl invariant. Therefore, the total variation of the matter action is
\begin{equation}
 \delta^W_\sigma S_{\rm m} = \int \underline{e} \, \sigma \left\{ \Theta^\mu{}_a e^a{}_\mu + \hat{\nabla}_\mu \mathfrak{D}^\mu \right\}\,,
\end{equation}
which gives rise to the N\"other identity for the dilation-current
\begin{equation}\label{dilation-cons-EW}
 \hat{\nabla}_\mu \mathfrak{D}^\mu = - \Theta^\mu{}_\mu\,.
\end{equation}
From the first equation in \eqref{improved-tensors}, we know that the trace of the improved energy-momentum tensor becomes
\begin{equation}
 \Theta^\mu{}_\mu = T^\mu{}_\mu + 2 \, \hat{\nabla}_\mu \Sigma^{\nu\mu}{}_\nu \, ,
\end{equation}
whereas the second one gives us the explicit form of the improved dilation-current. Putting all these formulae together, we can rewrite Eq.~\eqref{dilation-cons-EW} as
\begin{equation}
 \hat{\nabla}_\mu \Delta^\mu + 2 \hat{\nabla}_\mu \Sigma^\nu{}_\nu{}^\mu = - T^\mu{}_\mu - 2 \hat{\nabla}_\mu \Sigma^{\nu\mu}{}_\nu \,,
\end{equation}
which also implies the N\"other identity that was previously found in the Cartan formalism \eqref{trace-energy}. So, the two approaches yield the same result for Weyl invariance as well.

\subsection{Diffeomorphism invariance}\label{sect:diff-invariance-einstein-weyl}

Having established the N\"other identities which follow from Lorentz and Weyl gauge invariances, we finally turn our attention to those derived from diffeomorphism invariance. As we did for the Cartan formalism in the previous section, we are going to derive the diffeomorphisms conservation laws using the improved Einstein variations, i.e.~the covariant Lie derivatives defined in Sect.~\ref{sect:improved} and further explored in Appendix~\ref{algebra}. Given that we have split the spin-connection into torsion-free and torsionful parts, now $\tilde{\delta}^E_\xi$ takes the form
\begin{equation}
 \tilde{\delta}^E_\xi = \pounds_\xi + \delta^L_{\xi \cdot \hat{\omega}} + \delta^L_{\xi \cdot\hat{\Omega}} + \delta^W_{\xi \cdot S}\,,
\end{equation}
which includes the ``hatted'' connection.
Therefore, using the symmetry of the holonomic connection $\hat{\Gamma}_\mu = \mathring{\Gamma}_\mu + L_\mu$ and $\hat{\nabla}_\nu e^a{}_\mu = 0$, we can write the variations of the gravitational field variables as
\begin{equation}\label{Einstein-transformations-2}
\begin{split}
& \tilde{\delta}_\xi^E \, e^a{}_\mu  = e^a{}_\nu \hat{\nabla}_\mu \xi^\nu + \xi^\nu \hat{\Omega}^a{}_{b\nu} e^b{}_\mu
\,,\\
& \tilde{\delta}_\xi^E \, \hat{\Omega}^a{}_{b\mu}
= \xi^\nu \hat{\nabla}_\nu \hat{\Omega}^a{}_{b\mu} + \hat{\Omega}^a{}_{b\nu} \hat{\nabla}_\mu \xi^\nu + \xi^\nu [ \hat{\Omega}_\nu , \hat{\Omega}_\mu ]^a{}_b
\,,\\
& \tilde{\delta}_\xi^E \, S_\mu = \xi^\nu W_{\nu\mu}  \,.
\end{split}
\end{equation}
Exploiting the rules for the integration by parts discussed in Appendix \eqref{section-IntByParts}, the improved Einstein variation of the matter action with on-shell matter fields takes the form
\begin{equation}
 \begin{split}
  \tilde{\delta}^E_\xi S_{\rm m} = \int \underline{e} \, \xi^\nu \Bigl\{ -\hat{\nabla}_\mu \Theta^\mu{}_\nu + \Theta^\mu{}_\lambda \hat{K}^\lambda{}_{\mu\nu} + \Sigma^\mu{}_\lambda{}^\rho \hat{\nabla}_\nu \hat{K}^\lambda{}_{\rho\mu}  \\
   - \hat{\nabla}_\mu \bigl( \Sigma^\mu{}_\lambda{}^\rho \hat{K}^\lambda{}_{\rho\nu} \bigr) + \Sigma^\mu{}_\lambda{}^\rho [ \hat{K}_\nu, \hat{K}_\mu]^\lambda{}_\rho + \mathfrak{D}^\mu W_{\nu\mu} \Bigr\}\,.
 \end{split}
\end{equation}
The consequence is that the conservation law following from covariant diffeomorphism invariance is
\begin{equation}\label{energy-cons-WE1}
 \hat{\nabla}_\mu \Theta^\mu{}_\nu = \Theta^\mu{}_\lambda \hat{K}^\lambda{}_{\mu\nu} + \Sigma^\mu{}_\lambda{}^\rho \hat{\nabla}_\nu \hat{K}^\lambda{}_{\rho\mu} - \hat{\nabla}_\mu \left( \Sigma^\mu{}_\lambda{}^\rho \hat{K}^\lambda{}_{\rho\nu} \right) + \Sigma^\mu{}_\lambda{}^\rho [ \hat{K}_\nu, \hat{K}_\mu]^\lambda{}_\rho + \mathfrak{D}^\mu W_{\nu\mu}\,.
\end{equation}
Let us notice that, in the first term on the right hand side of the above equation, only the antisymmetric part of $\Theta_{[\mu\nu]}$ appears. We can then exploit the Eq.~\eqref{anti-sym-theta1} to write it in terms of the spin-density and the contortion tensor. A striking simplification occurs, for this term cancels exactly with the next-to-last in \eqref{energy-cons-WE1}. As a consequence, we can rewrite the N\"other identity \eqref{energy-cons-WE1}
\begin{equation}
\hat{\nabla}_\mu \Theta^\mu{}_\nu = \Sigma^\mu{}_\lambda{}^\rho \hat{\nabla}_\nu K^\lambda{}_{\rho\mu} - \hat{\nabla}_\mu \left( \Sigma^\mu{}_\lambda{}^\rho K^\lambda{}_{\rho\nu} \right) + \mathfrak{D}^\mu W_{\nu\mu}\,.
\end{equation}
The next step is provided by decomposing the energy-momentum tensor on the left hand side into its symmetric traceless, trace and antisymmetric irreducible components, using both Lorentz and Weyl invariance. Denoting the symmetric traceless part by $\theta^{\mu\nu}=\Theta^{\mu\nu}-\frac{1}{4}\Theta^\rho{}_\rho g^{\mu\nu}$, we have
\begin{equation}
\Theta^{\mu\nu} = \theta^{\mu\nu} - \frac{1}{4} g^{\mu\nu} \hat{\nabla}_\rho \mathfrak{D}^\rho + K^{\lambda\mu}{}_\rho \Sigma^{\rho\nu}{}_\lambda - K^{\lambda\nu}{}_\rho \Sigma^{\rho\mu}{}_\lambda\,.
\end{equation}
Thus, we can express the N\"other identity in terms as an equation for the divergence of the symmetric trace-free part of the energy-momentum tensor
\begin{align}\nonumber
\hat{\nabla}_\mu \theta^{\mu\nu} = & \, \hat{\nabla}_\mu \left( K^{\lambda\nu}{}_\rho \Sigma^{\rho\mu}{}_\lambda - K^{\lambda\mu}{}_\rho \Sigma^{\rho\nu}{}_\lambda \right) + \Sigma^{\mu\lambda\rho} \hat{\nabla}^\nu K_{\lambda\rho\mu} - \hat{\nabla}_\mu \left( \Sigma^{\mu\lambda\rho} K_{\lambda\rho}{}^\nu \right)\\
+ & \, \frac{1}{4} \hat{\nabla}^\nu \hat{\nabla}_\mu \mathfrak{D}^\mu + \mathfrak{D}_\mu W^{\nu\mu}
\end{align}
The special cases, studied in the previous section, of vanishing spin-current in Sect.~\ref{subsect:with-dilation-no-spin} or zero dilation-current in Sect.~\ref{subsect:with-spin-no-dilation}, can similarly be read off from the second and first line of the previous equation, respectively.

\subsubsection{The complete equivalence of Cartan-Weyl and Einstein-Weyl formalisms}

We have already observed that the Cartan-Weyl and Einstein-Weyl approaches yield equivalent N\"other identities for Lorentz and Weyl invariance. 
Now we aim at proving that the same goes for diffeomorphism-invariance as well. In order to do so, we need to derive the N\"other identity of Cartan's formalism \eqref{energy-cons} from Eq.~\eqref{energy-cons-WE1}.

We start by rewriting the left hand side of Eq.~\eqref{energy-cons-WE1}, using the first equation of \eqref{improved-tensors}
\begin{align}\label{EM-manip-step1}
 \hat{\nabla}_\mu \Theta^{\mu\nu} = & \, \hat{\nabla}_\mu T^{\mu\nu} + \hat{\nabla}_\mu \hat{\nabla}_\rho \left( \Sigma^{\mu\rho\nu} + \Sigma^{\nu\rho\mu} - \Sigma^{\rho\mu\nu} \right) \\\nonumber
 = & \, \tilde{\nabla}_\mu T^{\mu\nu} + \tilde{\tau}_\mu T^{\mu\nu} + \hat{K}_\lambda{}^\nu{}_\mu T^{\mu\lambda} + \frac{1}{2} \big[ \hat{\nabla}_\mu, \hat{\nabla}_\rho \big] \left( \Sigma^{\mu\rho\nu} + \Sigma^{\nu\rho\mu} - \Sigma^{\rho\mu\nu} \right) \, ,
\end{align}
where we have used the antisymmetry in the exchange of two indices, $\mu \leftrightarrow\rho $, in the expression in round parentheses in the first line. The first term on the right hand side of \eqref{energy-cons-WE1} can be combined with the part of the third one in which $\hat{\nabla}_\mu$ acts on the spin-current
\begin{align}\label{EM-manip-step2}
 \Theta^\mu{}_\lambda \hat{K}^\lambda{}_\mu{}^\nu - \hat{K}^\lambda{}_\rho{}^\nu \hat{\nabla}_\mu \Sigma^\mu{}_\lambda{}^\rho = & \, T^\mu{}_\lambda \hat{K}^\lambda{}_\mu{}^\nu + \hat{K}_{\lambda\mu}{}^\nu \hat{\nabla}_\rho \left( \Sigma^{\mu\rho\lambda} + \Sigma^{\lambda\rho\mu} - \Sigma^{\rho\mu\lambda} \right) - \hat{K}^\lambda{}_\rho{}^\nu \hat{\nabla}_\mu \Sigma^\mu{}_\lambda{}^\rho \\\nonumber
 = & \, T^\mu{}_\lambda \hat{K}^\lambda{}_{\mu\nu} \, .
\end{align}
In fact, the last two terms of the first line cancel each other, whereas the first two in the round parenthesis drop since the symmetrized indices are contracted with the antisymmetric ones of the contortion tensor. We note that the third term in the last line of \eqref{EM-manip-step1} combines with \eqref{EM-manip-step2} to give the required contraction of the energy-momentum tensor with the covariant torsion tensor.

Going back to the commutator of covariant derivatives in \eqref{EM-manip-step1}, after some algebra we find
\begin{align}\label{EM-manip-step3}
 \frac{1}{2} \big[ \hat{\nabla}_\mu, \hat{\nabla}_\rho \big] \left( \Sigma^{\mu\rho\nu} + \Sigma^{\nu\rho\mu} - \Sigma^{\rho\mu\nu} \right) = & \, \frac{1}{2} g^{\nu\sigma} \left\{ \hat{R}_{\sigma\lambda\rho\mu} \Sigma^{\lambda\rho\mu} - 4 W_{\rho\mu} \Sigma^{\mu\rho}{}_\sigma - 2 W_{\rho\mu} \Sigma_\sigma{}^{\rho\mu} \right\}\,.
\end{align}
Let us focus on the first term on the right hand side of the last equation. Using repeatedly the first Bianchi identity given in \eqref{Bianchi1-tor-free} for the ``hatted'' Riemann tensor, and exploit both Eq.~\eqref{Riem-hat-sym12} and the symmetry properties of the spin-current, yields
\begin{align}\label{EM-manip-step4}
 \hat{R}_{\sigma\lambda\rho\mu} \Sigma^{\lambda\rho\mu} = & \, - 2 \hat{R}_{\mu\lambda\sigma\rho} \Sigma^{\lambda\rho\mu} + 4 g_{\mu\lambda} W_{\sigma\rho} \Sigma^{\lambda\rho\mu} + 2 W_{\rho\mu} \Sigma_\sigma{}^{\rho\mu} \\\nonumber
 = & \, 2 \hat{R}_{\mu\rho\lambda\sigma} \Sigma^{\lambda\rho\mu} + 2 \hat{R}_{\mu\sigma\rho\lambda} \Sigma^{\lambda\rho\mu} + 4 W_{\sigma\rho} \Sigma^{\lambda\rho}{}_\lambda + 2 W_{\rho\mu} \Sigma_\sigma{}^{\rho\mu} \\\nonumber
 = & \, 2 \hat{R}_{\rho\mu\sigma\lambda} \Sigma^{\lambda\rho\mu} - 2 \hat{R}_{\sigma\mu\rho\lambda} \Sigma^{\lambda\rho\mu} + 4 g_{\mu\sigma} W_{\rho\lambda} \Sigma^{\lambda\rho\mu} + 4 W_{\sigma\rho} \Sigma^{\lambda\rho}{}_\lambda + 2 W_{\rho\mu } \Sigma_\sigma{}^{\rho\mu}\,.
\end{align}
The second term in the last line vanishes because of the first Bianchi identity, whereas the third and fifth are equal and opposite to those appearing in \eqref{EM-manip-step3}. Consequently, we have
\begin{equation}\label{EM-manip-step5}
 \frac{1}{2} \big[ \hat{\nabla}_\mu, \hat{\nabla}_\rho \big] \left( \Sigma^{\mu\rho\nu} + \Sigma^{\nu\rho\mu} - \Sigma^{\rho\mu\nu} \right) = \hat{R}^{\rho\lambda\nu}{}_\mu \Sigma^\mu{}_{\rho\lambda} + 2 W^\nu{}_\mu \Sigma^{\lambda\mu}{}_\lambda\,.
\end{equation}
Using the explicit form of the improved dilation-current \eqref{improved-tensors}, the last term in Eq.~\eqref{energy-cons-WE1} reads
\begin{equation}
 \mathfrak{D}_\mu W^{\nu\mu} = \Delta_\mu W^{\nu\mu} + 2 \Sigma^\lambda{}_{\lambda\mu} W^{\nu\mu} \, ,
\end{equation}
and the second term on the right hand side cancels out with the last term in \eqref{EM-manip-step5}.

All those terms which have not been considered up to now can be recognized straightforwardly as the torsional part of the full curvature tensor. Therefore, adding up all the contributions, we reobtain the N\"other identity that we have found in the Cartan formalism and repeate here for convenience
\begin{equation}
 \tilde{\nabla}_\mu T^\mu{}_\nu = - \tilde{\tau}_\mu T^\mu{}_\nu  + T^\mu{}_\rho \tilde{T}^\rho{}_{\nu\mu} + \Sigma^\mu{}_{\rho\lambda} \tilde{R}^{\rho\lambda}{}_{\nu\mu} + \Delta^\mu W_{\nu\mu} \,.
\end{equation}

\section{Conclusions}

We have given arguments to support the idea that Weyl symmetry
should be gauged in a gravitational theory with independent metric and connection
degrees of freedom. This is especially important if we assume that
local scale invariance is a fundamental symmetry of nature at high energies,
as it predicts the existence of a vector gauge potential for Weyl symmetry.
Our exploration has been mostly based on the fact that the independent connection
does not have to transform under Weyl rescalings as the Levi-Civita one, which is symmetric and metric compatible, does. In fact, it should not transform in that way, for example,
if we are trying to have a Weyl invariant notion of geodesic.

The path that we have followed is not dissimilar to the one of other authors,
and, in fact, we have achieved similar conclusions at some stages.
This can be seen in the construction that we have referred to as ``Cartan-Weyl'' formalism, which deals with a Weyl covariant torsion and a special form of nonmetricity that cancels exactly with the local gauging when the covariant derivative is gauged in the Weyl symmetry.
However, an important achievement of our analysis is that we have shown
how to define a generalization of torsion and contortion which is Weyl invariant,
rather than covariant, referred to as the ``Einstein-Weyl'' formalism. This has resulted in the definition of a new covariant derivative, $\hat{\nabla}$, which has several desirable features, besides the compatibility with the metric. In particular, this affine-connection induces a new splitting of the spin-connection $\omega^a{}_b=\hat{\omega}^a{}_b + \hat{\Omega}^a{}_b$, where
\begin{equation}
 \hat{\omega}^a{}_b= \frac{1}{2} \left( E_b \lfloor D e^a - E^a \lfloor D e_b + e_c \, E^a \lfloor E_b \lfloor De^c \right)
\end{equation}
is the torsion-free Weyl-invariant piece with affine Lorentz transformation, whereas $\hat{\Omega}^a{}_b$ is the Weyl-invariant torsion-full Lorentz tensor
(a formal definition of floor operator $\lfloor$ is given in Sect.\ \ref{sect:geom-noether}).
 One important property of $\hat{\nabla}$ is that we can integrate it by parts on scalars with Weyl weight $-4$ keeping Weyl invariance manifest.
We believe that this connection could be a staple
in future discussions of Weyl symmetry in the context of metric-affine theories of gravity.

One natural reason to incorporate Weyl symmetry in gravity and metric-affine theories is to try to construct
a theory that is ``complete'' above some ultraviolet scale, which is generally associated with the Planck mass.
In fact, metric and metric-affine theories are generally interpreted as effective
ones \cite{Baldazzi:2021kaf}, because, among other things, it is not clear if
they are predictive at all energy scales, especially if quantum mechanical effects are taken into account.
Needless to say, the formalism developed in this paper can also have potentially interesting implications
in the context of theories equivalent to general relativity, e.g.~\cite{Jimenez:2019woj,Jimenez:2021nup}, and it would be interesting to what geometrical model
a Weyl gauged theory could be equivalent to.

In the second part of the paper we have discussed how matter degrees of freedom should couple to the Weyl gauged geometry. In a completely general way,
we have derived the N\"other identities associated to Weyl, Lorentz and diffeomorphisms invariances. The identities constrain the currents that couple
the matter fields on-shell with the curvatures and covariant derivatives.

For the discussion of diffeomorphism invariance, we have shown how the use of
a generalization of the Lie derivative, occasionally known as covariant Lie derivative, is particularly convenient. Mathematically, the covariant Lie derivative allows us to maintain covariance under all gauge symmetries
at any moment of the computations, and results in a deformed version of
the algebra of diffeomorphisms in which parts associated with
local Weyl and Lorentz algebras are twisted by the presence of the connections.
The covariant Lie derivative generates what we referred to as ``improved'' diffeomorphisms, which we explored at length through the paper.

The prominent future perspective of our work would be to put into practice
the geometrical construction that we have pushed forward here. The Weyl gauging
potential $S_\mu$, which couples to the charges given by the weights of all fields,
can offer the opportunity to construct vector-tensor theories in which
the space of all parameters is heavily constrained by Weyl symmetry.
Since we have introduced a notion of Weyl invariant torsion, we are then free to
either include or exclude torsional degrees of freedom, depending on
the prescriptions that we want to follow when model-building. In this direction,
it would be interesting to find out which models with Weyl gauging are renormalizable and, eventually, asymptotically free, because they would pair well
with the remaining interactions of the standard model of particle physics.

More pragmatically, as we discussed at length in the main text,
a theory that is invariant under gauged Weyl transformations is scale invariant because the trace of its energy momentum tensor is a total divergence in the flat-space limit (of the so-called virial current, which couples to the Weyl gauge potential). The natural applications of the framework are thus systems in which scale or almost-scale invariances are realized naturally. These includes cosmological models of inflation, in which case the formalism has the potential to give an interpretation of scale invariance through fundamental fields among which torsion can play a pivotal role \cite{Karananas:2021gco}. Another important potential application is in the context of asymptotically safe gravity \cite{Reuter:1996cp}, which is an attempt to complete the (quantum) ultraviolet behavior of general relativity
through a nonperturbatively renormalizable scale invariant theory with gravitational degrees of freedom such as the metric and the connection.
We hope to elaborate more on these applications in future work.

\appendix



\section{Commutators of covariant derivatives and Bianchi identities}\label{sect::AppendixCommutators}

In this appendix we give some relevant identities for the two covariant derivatives used in the main text, $\tilde{\nabla}$ and $\hat{\nabla}$, introduced in Sects.~\ref{sect:take-one} and \ref{sect:take-two}. For both of them, we give the commutators of two covariant derivatives, we consider the nontrivial contractions of the curvature tensors, and we write down the appropriate generalizations of the two Bianchi identities.

\subsection*{The symmetric Weyl invariant connection}

We begin with the torsion-free connection $\hat{\nabla}$. Given a vector field $v^\mu$ with Weyl weight $w_v$, the commutator of two covariant derivatives is
\begin{equation}\label{comm-rel-torsionfree}
 \bigl[ \hat{\nabla}_\rho, \hat{\nabla}_\mu \bigr] v^\lambda = \hat{R}^\lambda{}_{\sigma\rho\mu} v^\sigma + w_v W_{\rho\mu} v^\lambda,
\end{equation}
where $W_{\rho\mu}=\partial_\rho S_\mu - \partial_\mu S_\rho$ is the Weyl $2$-form, and $\hat{R}^\lambda{}_{\sigma\rho\mu}$ is the Weyl invariant curvature tensor constructed from the full connection $\hat{\Gamma}_\mu = \mathring{\Gamma}_\mu + L_\mu$. In terms of the curvature of the metric compatible symmetric connection, we have
\begin{align}
 \hat{R}^\lambda{}_{\sigma\rho\mu} & = \mathring{R}^\lambda{}_{\sigma\rho\mu} + \mathring{\nabla}_\rho L^\lambda{}_{\sigma\mu} - \mathring{\nabla}_\mu L^\lambda{}_{\sigma\rho} + \bigl[ L_\rho, L_\mu \bigr]^\lambda{}_\sigma\\\nonumber
 & = \mathring{R}^\lambda{}_{\sigma\rho\mu} + \delta^\lambda{}_\sigma W_{\rho\mu} + \delta^\lambda{}_\mu \mathring{\nabla}_\rho S_\sigma - \delta^\lambda{}_\rho \mathring{\nabla}_\mu S_\sigma - g_{\sigma\mu} \mathring{\nabla}_\rho S^\lambda + g_{\sigma\rho} \mathring{\nabla}_\mu S^\lambda\\\nonumber
 & + \delta^\lambda{}_\rho \left( S_\mu S_\sigma - g_{\mu\sigma} S^2 \right) - \delta^\lambda{}_\mu \left( S_\rho S_\sigma - g_{\rho\sigma} S^2 \right) + g_{\mu\sigma} S_\rho  S^\lambda - g_{\rho\sigma} S_\mu  S^\lambda\,.
\end{align}
Notice that the full curvature tensor does \emph{not} have definite symmetry properties. Indeed, even though it is antisymmetric in the pair $(\rho ,\mu)$, it contains both symmetric and antisymmetric terms in the pair $(\lambda,\sigma)$. However, $\delta^\lambda{}_\sigma W_{\rho\mu}$ is the only term symmetric under the exchange of $(\lambda,\sigma)$, in fact it is a pure trace and, clearly, Weyl invariant.
Therefore, since the whole expression is Weyl invariant, $\hat{R}^\lambda{}_{\sigma\rho\mu} - \delta^\lambda{}_\sigma W_{\rho\mu}$ is also a Weyl invariant tensor, which is antisymmetric in $(\lambda,\sigma)$. We notice that under exchange of the first two indices the curvature satisfies
\begin{equation}\label{Riem-hat-sym12}
 \hat{R}^{\lambda\rho}{}_{\mu\nu} = 2 g^{\lambda\rho} W_{\mu\nu} - \hat{R}^{\rho\lambda}{}_{\mu\nu}\,.
\end{equation}

We have three nonvanishing contractions of the curvature tensor
\begin{subequations}\label{ContractionsAndRicci}
 \begin{align}
  & \hat{R}^\lambda{}_{\lambda\rho\mu}  = 4 W_{\rho\mu} \,, \\
  & \hat{R}^\lambda{}_{\sigma\lambda\mu}  \equiv \widehat{Ric}_{\sigma\mu} = \mathring{R}_{\sigma\mu} + W_{\sigma\mu} - g_{\sigma\mu} \mathring{\nabla}_\rho S^\rho - 2 \mathring{\nabla}_\mu S_\sigma + 2 S_\mu S_\sigma - 2 g_{\mu\sigma} S^2 \,, \\
  & \hat{R}^\lambda{}_{\sigma\rho\mu} g^{\sigma\mu} g_{\lambda\nu} = \widehat{Ric}_{\nu\rho} - 2 W_{\nu\rho} \,.
 \end{align}
\end{subequations}
The first contraction \cite{charap1974gauge}, which is known in the literature with the name of homothetic curvature tensor, is nonvanishing as it often happens with theories in which the local symmetry group is enhanced (e.g., to $GL(4)$) \cite{Vazirian:2013baa}. In our geometry the homothetic curvature is proportional to the Weyl curvature $2$-form, which is the field strength of local scale transformations.

We also notice that the last contraction in \eqref{ContractionsAndRicci} is a linear superposition of the previous two, so we concentrate on the second. Whereas the first contraction is clearly antisymmetric in the two indices, the second one is still reducible. To highlight this feature, we give the explicit form of its symmetric and antisymmetric parts
\begin{subequations}\label{sym-asym-Ricci}
 \begin{align}
  \widehat{Ric}_{(\mu\sigma)} & = \mathring{R}_{\mu\sigma} - \left( \mathring{\nabla}_\mu S_\sigma + \mathring{\nabla}_\sigma S_\mu + g_{\mu\sigma} \mathring{\nabla}_\rho S^\rho \right) + 2 \left( S_\mu S_\sigma - g_{\mu\sigma} S^2 \right) \,, \\
  \widehat{Ric}_{[\mu\sigma]} & = 2 W_{\mu\sigma}\,.
 \end{align}
\end{subequations}
Only the symmetric part contributes to the trace that results in the Weyl covariant scalar curvature
\begin{equation}
 \hat{R} = \mathring{R} - 6 \mathring{\nabla}_\rho S^\rho - 6 S^2\,.
\end{equation}
As a consistency check, we note that if we take the Weyl gauge potential to be a pure gauge of the form $S_\mu = \partial_\mu \omega$, the previous equation gives us the transformation rule of the Levi-Civita scalar curvature, $\mathring{R}$, when we apply a standard Weyl transformation $g_{\mu\nu} \rightarrow e^{2 \omega(x)} g_{\mu\nu}$ to the metric.

Since the affine connection $\hat{\Gamma}$ is symmetric, we expect that the first Bianchi identities follow straightforwardly. In fact, it becomes
\begin{align}\label{Bianchi1-tor-free}
 \hat{R}^\lambda{}_{[\sigma\mu\nu]} = \delta^\lambda{}_{[\sigma} W_{\mu\nu]} + \delta^\lambda{}_{[\nu} W_{\mu\sigma]} = 0\,.
\end{align}
Using again the symmetry and of the affine connection as well as the Weyl invariance of both curvature tensors, we obtain two more Bianchi identities for the curvature $2$-forms of the Weyl geometry
\begin{subequations}
 \begin{align}
  \hat{\nabla}_{[\mu} \hat{R}^\lambda{}_{|\sigma|\nu\rho]} = 0 \,,\\
  \hat{\nabla}_{[\mu} W_{\nu\rho]} = 0\,.
 \end{align}
\end{subequations}
Exploiting the second Bianchi identities for the curvature tensor, the contractions \eqref{ContractionsAndRicci} and the second equation in \eqref{sym-asym-Ricci}, we obtain the proper generalization of the contracted Bianchi identities
\begin{equation}\label{contr-Bianchi-tor-free}
 \hat{\nabla}^\mu \left( \widehat{Ric}_{\mu\rho} - W_{\mu\rho} - \frac{1}{2} g_{\mu\rho} \hat{R} \right) = 0\,.
\end{equation}
Notice that, taking the symmetric and antisymmetric parts of the Ricci tensor, we can rewrite the previous equation as  
\begin{equation}
 \hat{\nabla}^\mu \left( \widehat{Ric}_{(\mu\rho)} - \frac{1}{2} g_{\mu\rho} \hat{R} \right) =  - \hat{\nabla}^\mu W_{\mu\rho} \, .
\end{equation}

\subsection*{The nonsymmetric torsionful connection}

The commutator of two covariant derivatives on a vector field $v^\rho$ with weight $w_v$ is
\begin{equation}
\bigl[ \tilde{\nabla}_\mu,\tilde{\nabla}_\nu \bigr] v^\rho
=
\tilde{R}^\rho{}_{\sigma\mu\nu} v^\sigma
+ \tilde{T}^\sigma{}_{\nu\mu} \tilde{\nabla}_\sigma v^\rho
+ w_v W_{\mu\nu} v^\rho
\,,
\end{equation}
where the Weyl invariant curvature tensor of $\tilde{\nabla}$ is
\begin{align}
 \tilde{R}^\rho{}_{\lambda\mu\nu} = &
 \, \hat{R}^\rho{}_{\lambda\mu\nu} + \hat{\nabla}_\mu \hat{K}^\rho{}_{\lambda\nu} - \hat{\nabla}_\nu \hat{K}^\rho{}_{\lambda\mu} + \big[ \hat{K}_\mu, \hat{K}_\nu \big]^\rho{}_\lambda \\
 \nonumber
 = & \, \hat{R}^\rho{}_{\lambda\mu\nu} + \big[ \hat{K}_\mu, \hat{K}_\nu \big]^\rho{}_\lambda + \mathring{\nabla}_\mu \hat{K}^\rho{}_{\lambda\nu} - \mathring{\nabla}_\nu \hat{K}^\rho{}_{\lambda\mu}
  + S_\lambda \tilde{T}^\rho{}_{\mu\nu} + S^\kappa \left( g_{\lambda\mu} \hat{K}^\rho{}_{\kappa\nu} - g_{\lambda\nu} \hat{K}^\rho{}_{\kappa\mu} \right)
  \\\nonumber
  & \, + \left( \delta^\rho{}_\mu S_\kappa - g_{\kappa\mu} S^\rho \right) \hat{K}^\kappa{}_{\lambda\nu} - \left( \delta^\rho{}_\nu S_\kappa - g_{\kappa\nu} S^\rho \right) \hat{K}^\kappa{}_{\lambda\mu} \\\nonumber
 = & \, \mathring{R}^\rho{}_{\lambda\mu\nu} + \delta^\rho{}_\lambda W_{\mu\nu} + \delta^\rho{}_\nu \mathring{\nabla}_\mu S_\lambda - \delta^\rho{}_\mu \mathring{\nabla}_\nu S_\lambda + g_{\lambda\mu} \mathring{\nabla}_\nu S^\rho - g_{\lambda\nu} \mathring{\nabla}_\mu S^\rho\\\nonumber
  & \, + \delta^\rho{}_\mu \left( S_\lambda S_\nu - g_{\lambda\nu} S^2 \right) - \delta^\rho{}_\nu \left( S_\lambda S_\mu - g_{\lambda\mu} S^2 \right) + g_{\lambda\nu} S_\mu S^\rho - g_{\lambda\mu} S_\nu S^\rho\\\nonumber
  & \, + \big[ \hat{K}_\mu, \hat{K}_\nu \big]^\rho{}_\lambda + \mathring{\nabla}_\mu \hat{K}^\rho{}_{\lambda\nu} - \mathring{\nabla}_\nu \hat{K}^\rho{}_{\lambda\mu} + \delta^\rho{}_\mu S_\kappa \hat{K}^\kappa{}_{\lambda\nu} - \delta^\rho{}_\nu S_\kappa \hat{K}^\kappa{}_{\lambda\mu} \\\nonumber
  & \, + g_{\mu\lambda} S^\kappa \hat{K}^\rho{}_{\kappa\nu} - g_{\nu\lambda} S^\kappa \hat{K}^\rho{}_{\kappa\mu} + S_\lambda \tilde{T}^\rho{}_{\mu\nu} - S^\rho \tilde{T}_{\lambda\mu\nu} \, ,
\end{align}
where we have also used the results of the previous subsection to expand the ``hatted'' Riemann tensor, as well as the explicit form of the covariant contortion-tensor. The commutation relations have the same form when applied to a Lorentz vector. For example, if we take $v^a = e^a{}_\rho v^\rho$, then we have
\begin{equation}
 \bigl[ \tilde{\nabla}_\mu,\tilde{\nabla}_\nu \bigr] v^a
 =
 \tilde{R}^a{}_{b\mu\nu} v^b
 + \tilde{T}^\lambda{}_{\nu\mu} \tilde{\nabla}_\lambda v^a
 + (w_v - 1) W_{\mu\nu} v^a
 \,,
\end{equation}
where $ \tilde{R}^a{}_{b\mu\nu} 
=e^a{}_\rho E^\sigma{}_b \tilde{R}^\rho{}_{\sigma\mu\nu} $ are the components
of the curvature $2$-form, $\tilde{{\cal R}}^a{}_b=\frac{1}{2}\tilde{R}^a{}_{b\mu\nu} dx^\mu\wedge dx^\nu$,
$W_{\mu\nu}= \partial_\mu S_\nu-\partial_\nu S_\mu$ are the components of the Weyl $2$-form,
${\cal W}=dS=\frac{1}{2}W_{\mu\nu} dx^\mu\wedge dx^\nu$,
and $\tilde{T}^\sigma{}_{\mu\nu}=E^\sigma{}_a \tilde{T}^a{}_{\mu\nu}$ those 
of the torsion $2$-form, $ \tilde{\mathcal{T}}^a = \tilde{D} e^a$.
Notice that the weight of the tensor depends
on the position (upper or lower) of its holonomic indices,
because both the tetrad and the metric have nonzero Weyl weight.
For example, if we define $z_\mu = g_{\mu\nu} v^\nu$, then $w_z=w_v + 2$ and
\begin{equation}
\bigl[ \tilde{\nabla}_\mu,\tilde{\nabla}_\nu \bigr] z_\rho
=
-\tilde{R}^\sigma{}_{\rho\mu\nu} z_\sigma
+ \tilde{T}^\sigma{}_{\nu\mu} \tilde{\nabla}_\sigma z^\rho
+ (w_v + 2) W_{\mu\nu} z^\rho
\,.
\end{equation}

Now we turn our attention to the nonvanishing contractions of the full curvature tensor, exploiting the definition of the torsion-vector \eqref{torsion-trace} to simplify the expression. There are three, but only two are truly independent
\begin{subequations}\label{ContractionsAndRicci-torsion}
 \begin{align}
 \tilde{R}^\rho{}_{\rho\mu\nu} = & 4 W_{\mu\nu} \,, \\
 \tilde{R}^\mu{}_{\lambda\mu\nu}  \equiv & \widetilde{Ric}_{\lambda\nu} = \mathring{R}_{\lambda\nu}
  - g_{\lambda\nu} \mathring{\nabla}_\mu S^\mu + W_{\lambda\nu} - 2 \mathring{\nabla}_\nu S_\lambda + 2 S_\lambda S_\nu - 2 g_{\lambda\nu} S^2\\\nonumber
 & + \mathring{\nabla}_\mu \hat{K}^\mu{}_{\lambda\nu} + \mathring{\nabla}_\nu \tilde{\tau}_\lambda - \tilde{\tau}_\mu \hat{K}^\mu{}_{\lambda\nu} - \hat{K}^\mu{}_{\rho\nu} \hat{K}^\rho{}_{\lambda\mu}\\\nonumber
 & +  S_\mu \left( \tilde{T}^\mu{}_{\nu\lambda} + \tilde{T}_\nu{}^\mu{}_\lambda \right) - S_\lambda \tilde{\tau}_\nu + g_{\lambda\nu} S_\mu \tilde{\tau}^\mu
 \,, \\
 \tilde{R}^\rho{}_{\lambda\mu\nu} g^{\lambda\nu} g_{\rho\sigma} = & \widetilde{Ric}_{\sigma\mu} - 2 W_{\sigma\mu}\,,
 \end{align}
\end{subequations}
but, in analogy with the torsionless case, only two are truly independent.
As in the previous case, we give the expressions of the symmetric and antisymmetric parts of the Ricci tensor
\begin{subequations}\label{sym-asym-Ricci-torsion}
 \begin{align}
 \widetilde{Ric}_{(\lambda\nu)}  = &\,  \mathring{R}_{\lambda\nu} - \left( \mathring{\nabla}_\lambda S_\nu + \mathring{\nabla}_\nu S_\lambda + g_{\lambda\nu} \mathring{\nabla}_\rho S^\rho \right) + 2 \left( S_\lambda S_\nu - g_{\lambda\nu} S^2 \right)\\\nonumber
 & + \, \frac{1}{2} \left( \mathring{\nabla}_\mu  + S_\mu - \tilde{\tau}_\mu \right) \left( \tilde{T}_\nu{}^\mu{}_\lambda + \tilde{T}_\lambda{}^\mu{}_\nu \right) \\\nonumber
 & + \, \frac{1}{2} \left( \mathring{\nabla}_\nu \tilde{\tau}_\lambda + \mathring{\nabla}_\lambda \tilde{\tau}_\nu - S_\nu \tilde{\tau}_\lambda - S_\lambda \tilde{\tau}_\nu \right) + g_{\lambda\nu} S_\mu \tilde{\tau}^\mu\\\nonumber
 & + \, \frac{1}{4} \left( \tilde{T}_\lambda{}^{\mu\rho} \tilde{T}_{\nu\mu\rho} - \tilde{T}_\lambda{}^\rho{}_\mu \tilde{T}_\rho{}^\mu{}_\nu - \tilde{T}_\nu{}^\rho{}_\mu \tilde{T}_\rho{}^\mu{}_\lambda \right) \,, \\
 \widetilde{Ric}_{[\lambda\nu]} = & \, 2 W_{\lambda\nu} - \frac{1}{2} \left( \mathring{\nabla}_\mu - \tilde{\tau}_\mu + 2 S_\mu \right) \tilde{T}^\mu{}_{\lambda\nu} + \frac{1}{2} S_\mu \left( \tilde{T}_\nu{}^\mu{}_\lambda - \tilde{T}_\lambda{}^\mu{}_\nu \right) \\\nonumber
 & + \, \frac{1}{2} \left( \mathring{\nabla}_\nu \tilde{\tau}_\lambda - \mathring{\nabla}_\lambda \tilde{\tau}_\nu + S_\nu \tilde{\tau}_\lambda - S_\lambda \tilde{\tau}_\nu  \right) + \frac{1}{4} \left( \tilde{T}_\lambda{}^\mu{}_\rho \tilde{T}^\rho{}_{\mu\nu} - \tilde{T}_\nu{}^\mu{}_\rho \tilde{T}^\rho{}_{\mu\lambda} \right) \,.
 \end{align}
\end{subequations}
The Weyl covariant scalar curvature in presence of torsion is thus
\begin{equation}
 \tilde{R} = \mathring{R} - 6 \mathring{\nabla}_\rho S^\rho - 6 S^2 + 2 \mathring{\nabla}_\mu \tilde{T}^{\nu\mu}{}_\nu - \tilde{T}^\mu{}_{\rho\mu} \tilde{T}^{\nu\rho}{}_\nu + \frac{1}{4} \tilde{T}^{\mu\nu\rho} \tilde{T}_{\mu\nu\rho} + \frac{1}{2} \tilde{T}^{\mu\nu\rho} \tilde{T}_{\rho\nu\mu} + 4 S_\nu \tilde{T}^{\mu\nu}{}_\mu \,.
\end{equation}
We can use the relation $\hat{\nabla}_\mu \tilde{T}^{\nu\mu}{}_\nu= \mathring{\nabla}_\mu \tilde{T}^{\nu\mu}{}_\nu + 2 S_\mu \tilde{T}^{\nu\mu}{}_\nu$, which allows to write the scalar curvature in a manifestly Weyl covariant way
\begin{equation}
 \tilde{R} = \hat{R} + 2 \hat{\nabla}_\mu \tilde{T}^{\nu\mu}{}_\nu - \tilde{T}^\mu{}_{\rho\mu} \tilde{T}^{\nu\rho}{}_\nu + \frac{1}{4} \tilde{T}^{\mu\nu\rho} \tilde{T}_{\mu\nu\rho} + \frac{1}{2} \tilde{T}^{\mu\nu\rho} \tilde{T}_{\rho\nu\mu}
 \,.
\end{equation}

The Bianchi identities for the connection $\tilde{\nabla}$ with both torsion and Weyl gauging are slightly more complicated, mostly because of the presence of the Weyl $2$-form $W_{\mu\nu}$.
We find
\begin{eqnarray}\label{Bianchi1-tor}
 \, \tilde{\nabla}_{[\mu} \tilde{T}^\lambda{}_{\nu\rho]}
 &=&
 \tilde{R}^\lambda{}_{[\mu\nu\rho]} + \tilde{T}^\kappa{}_{[\mu\nu} \tilde{T}^\lambda{}_{\rho]\kappa}\,, \\ \label{Bianchi2-tor}
 \, \tilde{\nabla}_{[\mu} \tilde{R}^\lambda{}_{|\sigma|\nu\rho]}
 &=&
 \tilde{T}^\kappa{}_{ [\mu \nu} R^\lambda{}_{|\sigma|\rho] \kappa}
 \,,
 \\
 \, \tilde{\nabla}_{[\mu} W_{\nu\rho]}
 &=&
 \tilde{T}^\lambda{}_{ [\mu \nu} W_{\rho] \lambda} \,.
\end{eqnarray}
The first equation could be derived by expressing the torsion tensors on the left hand side as differences of contortion tensors and exploiting $\hat{R}^\lambda{}_{[\mu\nu\rho]}=0$. Nevertheless, the most efficient way of deriving all the above relations is using Cartan's differential-form formalism, in which they read
\begin{eqnarray}
\tilde{D} \, \tilde{\cal T}^a
&=&
\tilde{ \cal R}^a{}_b \wedge e^b +{\cal W} \wedge e^a \,,\\
\tilde{D} \, \tilde{ \cal R}^a{}_b
&=&
0 \,,
\\
\tilde{D} \, {\cal W}
&=&
0 \,.
\end{eqnarray}
The only caveat for passing from the latter set of equations to the former is that $\tilde{D}E^\lambda{}_a$ does not vanish, which complicates slightly the derivation of the first relation \eqref{Bianchi1-tor}.

Notice that, since the torsion $2$-form $\tilde{\mathcal{T}}^a$ has Weyl weight $w_{\tilde{\mathcal{T}}^a}=1$, the Weyl $2$-form appears on the right hand side of the corresponding Bianchi identity. On the other hand, the holonomic torsion tensor, as well as the Riemann and Weyl curvatures, are Weyl invariant, and this is the reason why no such contributions appears on the right hand side of \eqref{Bianchi1-tor}. Let us finally remark that passing from the differential form to the holonomic formalism results in new terms on the right hand side, in which the curvature tensors are contracted with the torsion itself.

Contracting two pairs of indices in \eqref{Bianchi2-tor}, we obtain the contracted Bianchi identities for the curvature tensor
\begin{equation}
 \tilde{\nabla}_\mu \tilde{R} - 2 \tilde{\nabla}^\nu \widetilde{Ric}_{\nu\mu} + 2 \tilde{\nabla}^\nu W_{\nu\mu} = 2 \tilde{T}^{\rho\nu}{}_\mu \left( \widetilde{Ric}_{\nu\rho} - W_{\nu\rho} \right) + \tilde{T}^\rho{}_{\nu\sigma} \tilde{R}^{\nu\sigma}{}_{\mu\rho}\,.
\end{equation}

Since we now have another nontrivial differential Bianchi identity \eqref{Bianchi1-tor}, we also have one more contracted identity, which can be derived from \eqref{Bianchi1-tor} and by contracting the upper index with any of the lower ones. Denoting the torsion vector by $\tilde{\tau}_\mu \equiv \tilde{T}^\nu{}_{\mu\nu}$ (as in the main text), the identity becomes
\begin{equation}
 ( \tilde{\nabla} + \tilde{\tau} )_\nu \left( \tilde{T}^\nu{}_{\mu\rho} + \delta^\nu{}_\mu \tilde{\tau}_\rho - \delta^\nu{}_\rho \tilde{\tau}_\mu \right) = 2 W_{\mu\rho} \,,
\end{equation}
where we recognize the modified torsion tensor on the left hand side.

A brief comment on the coupling with fermionic fields before closing in with the appendix. In the Einstein-Cartan formalism, the modified torsion-tensor is ``sourced'' by the spin-density of fermionic fields (see, for example, the Sciama-Kibble field equations in \cite{Trautman:2006fp}). In the case of Dirac fields, it has only vector and axial-vector irreducible components (see, for example, \cite{Freidel:2005sn,gasperini2013theory,Karananas:2021zkl}). Furthermore, it is also well-known that the Weyl potential decouples from the Dirac Lagrangian \cite{Oda:2020yyv}. On the other hand, in a Weyl gauged analog of Einstein-Cartan theory, the algebraic equations of motion for the torsion tensor read
\begin{equation}
\phi^2 \left( \tilde{T}^\nu{}_{\mu\rho} + \delta^\nu{}_\mu \tilde{\tau}_\rho - \delta^\nu{}_\rho \tilde{\tau}_\mu \right) = a \Sigma^\nu{}_{\mu\rho} \, ,
\end{equation}
with some unspecified numerical factor $a$. Applying $(\tilde{\nabla} + \tilde{\tau})_\nu$ on both sides of the previous equation and exploiting Eq.~\eqref{anti-sym-energy}, we obtain the antisymmetric part of the energy-momentum tensor in a Weyl gauged Einstein-Cartan theory
\begin{equation}
 a T_{[\mu\rho]} = 2 \phi^2 W_{\mu\rho} + \left( \tilde{T}^\nu{}_{\mu\rho} + \delta^\nu{}_\mu \tilde{\tau}_\rho - \delta^\nu{}_\rho \tilde{\tau}_\mu \right) (\hat{\nabla} + \tilde{\tau})_\nu \phi^2 \,.
\end{equation}


\section{Integration by parts}\label{section-IntByParts}

The derivations of the N\"other identites in the main text require multiple
uses of the integration by parts of independent connections, which is not as straightforward as with the unique symmetric metric compatible connection $\mathring{\nabla}$.

Mathematically speaking, integration on a $d$-dimensional manifold $\mathcal{M}$ is defined by appropriately gluing together the integration of a $d$-form over local charts with the aid of the partition of the unity. In this context, integration by parts is simply Stokes' theorem. Given a $(d-1)$-form $\zeta$, we have $\int_{\mathcal{M}} d \zeta = \int_{\partial \mathcal{M}} \zeta$, up to an orientation dependent sign.
Therefore, requiring that $\zeta$ vanish on the boundary of the manifold (for example at spatial infinity in general relativity),
Stokes' theorem becomes $\int_{\mathcal{M}} d\zeta = 0$, and we can safely integrate by parts. If the boundary does not exist, $\partial \mathcal{M} = \emptyset$, the same result holds. For simplicity, we assume that we can cover
our manifold $\mathcal{M}$ with a unique coordinate chart and that the coordinates range on the entire real axis (the generalization follows straightforwardly using
the standard tools to prove Stokes' theorem).

Physically speaking, we need a volume form, generally chosen $\underline{e}=\sqrt{-g}$, and integration must be defined over scalar densities.
When considering the connection $\mathring{\nabla}$ and the explicit form of its components $\mathring{\Gamma}^\mu{}_{\nu\mu}$, it is trivial to show
\begin{equation}\label{cov-divergence}
 \mathring{\nabla}_\mu v^\mu = \frac{1}{\sqrt{-g}} \partial_\mu \left( \sqrt{-g} v^\mu \right) \,.
\end{equation}
The consequence is that the volume form simplifies when we integrate and are left
with a partial derivative
\begin{equation}
 \int \sqrt{-g} \, \mathring{\nabla}_\mu v^\mu = \int \partial_\mu \left( \sqrt{-g} v^\mu \right) dx^1 \cdots dx^d = \sum_\mu \int \left( \sqrt{-g} v^\mu \right) \bigg|^\infty_{-\infty} dx^1 \cdots \widehat{dx^\mu} \cdots dx^d\,,
\end{equation}
where the notation indicates with a hat that $dx^\mu$ is excluded from the final expression.
With hindsight, in order to integrate by parts, we need our covariant differential operator to display the crucial property \eqref{cov-divergence}.

Now we take into account the ``hatted'' covariant derivative defined in Eq.~\eqref{eq:nablahat-definition}. The trace of the distortion tensor is $L^\mu{}_{\nu\mu}= 4 S_\nu$ (the numerical factor comes from the dimension of spacetime $d=4$). Considering a vector field $v^\mu$ with Weyl weight $w_v$ and using \eqref{cov-divergence}, we expand its covariant divergence as
\begin{equation}
 \hat{\nabla}_\mu v^\mu = \frac{1}{\sqrt{-g}} \partial_\mu \left( \sqrt{-g} v^\mu \right) + (4 + w_v) S_\mu v^\mu = \frac{1}{\sqrt{-g}} \hat{D}_\mu \left(\sqrt{-g}v^\mu\right),
\end{equation} 
where in the last step we used the fact that $w(\sqrt{-g})=4$ (again, the dimension of spacetime $d=4$), and $\hat{D}$ is the gauge-covariant derivative (acting only as gauge derivative, but blind to coordinate indices). We clearly see that integration by parts with respect to the ``hatted'' differential operator can be safely performed \emph{only} for vector fields $v^\mu$ with Weyl weight $w_v=-4$, that is, when $\hat{D}_\mu \left(\sqrt{-g}v^\mu\right) = \partial_\mu \left(\sqrt{-g}v^\mu\right)$. Fortunately, this is the case in all the relevant calculations carried out in the main text.

It is important to realize that, contrary to $\hat{\nabla}$, the complete covariant derivative $\tilde{\nabla}$ does not possess the crucial property \eqref{cov-divergence} which is needed for a straightforward integration by parts. Indeed, considering the covariant divergence of a vector field with Weyl weight $w_v$
\begin{equation}
 \tilde{\nabla}_\mu v^\mu = \hat{\nabla}_\mu v^\mu + \hat{K}^\mu{}_{\nu\mu} v^\nu = \frac{1}{\sqrt{-g}} \hat{D}_\mu \left( \sqrt{-g} v^\mu \right) + \tilde{T}^\mu{}_{\mu\nu} v^\nu\,.
\end{equation}
We see that the last term, proportional to the torsion vector, cannot be written in the form \eqref{cov-divergence} and, in general, does not vanish. Consequently, we cannot use the full covariant derivative $\tilde{\nabla}$ to integrate by parts. This is the reason why, throughout the whole paper, we have always used either $\mathring{\nabla}$ or $\hat{\nabla}$ to integrate by parts. It is most convenient to use $\hat{\nabla}$ instead of $\mathring{\nabla}$, for the former maintains Weyl covariance manifest.

Notwithstanding the ``bad'' behavior of $\tilde{\nabla}$, we might be interested in finding the general rule for appropriately integrating it by parts. To work it out, let us consider a vector-field $v^\nu$ and a tensor field $z^\mu{}_\nu$, such that the sum of their Weyl weights is $w_v + w_z = -4$. Expressing schematically $\tilde{\nabla}_\mu = \hat{\nabla}_\mu + \hat{K}_\mu $, we find
\begin{align}
 \int \sqrt{-g} v^\nu \tilde{\nabla}_\mu z^\mu{}_\nu & = \int \sqrt{-g} v^\nu \hat{\nabla}_\mu z^\mu{}_\nu + \int \sqrt{-g} v^\nu \left( \hat{K}^\mu{}_{\lambda\mu} z^\lambda{}_\nu - \hat{K}^\lambda{}_{\nu\mu} z^\mu{}_\lambda \right)\\\nonumber
 & = - \int \sqrt{-g} z^\mu{}_\nu \hat{\nabla}_\mu v^\nu + \int \sqrt{-g} z^\mu{}_\nu \left( \hat{K}^\lambda{}_{\mu\lambda} v^\nu - \hat{K}^\nu{}_{\lambda\mu} v^\lambda \right)\\\nonumber
 & = - \int \sqrt{-g} z^\mu{}_\nu \tilde{\nabla}_\mu v^\nu + \int \sqrt{-g} z^\mu{}_\nu \hat{K}^\lambda{}_{\mu\lambda} v^\nu \,.
\end{align}
The result that we have just found is valid even if we trade the holonomic index $\nu$ for an arbitrary collection of Latin and Greek indices. To highlight this feature, we are going to label this general set of indices with a multi-index $I$. Therefore, we have the rule
\begin{equation}
 \int \sqrt{-g} v^I \tilde{\nabla}_\mu z^\mu{}_I = - \int \sqrt{-g} z^\mu{}_I \tilde{\nabla}_\mu v^I + \int \sqrt{-g} v^I z^\mu{}_I \hat{K}^\lambda{}_{\mu\lambda} \,.
\end{equation}
Using the fact that $\hat{K}^\lambda{}_{\mu\lambda} = - \tilde{\tau}_\mu$ (see Eq.~ \eqref{torsion-trace}), we can rewrite the previous equation compactly
\begin{equation}
\int \sqrt{-g} v^I (\tilde{\nabla} + \tilde{\tau})_\mu z^\mu{}_I = - \int \sqrt{-g} z^\mu{}_I \tilde{\nabla}_\mu v^I\,,
\end{equation}
which thus becomes the go-to formula for integration by parts of $\tilde{\nabla}$
(recall that $w_v + w_z = -4$).


\section{Covariant Lie derivatives and the extended algebra}\label{algebra}

In the main text, specifically Sects.~\ref{sect:improved} and \ref{sect:diff-invariance-einstein-weyl}, we have defined and applied a covariant generalization of the Lie derivative, $\widetilde{\pounds}_\xi$. Such extension goes under the name of covariant Lie derivative and can be found, \emph{mutatis mutandis}, in the  literature of metric-affine gravity (see, for example, \cite{Gronwald:1997jd}).
In the present paper, $\widetilde{\pounds}_\xi$ provides an ``improvement''
to the standard infinitesimal diffeomorphisms, $\tilde{\delta}^E_{\xi}=\widetilde{\pounds}_\xi$, which is covariant under all gauge groups. We use the two symbols interchangeably.
Since its application to Lorentz-Weyl gauge theories has never been carried out, at least to our knowledge, we prove some of the main properties in this appendix.

Before diving into the details of the algebra generated by $\widetilde{\pounds}$, it is important to stress the main differences between the covariant Lie derivative and (any) ordinary covariant derivative. The first observation is that we can give meaning to the covariant Lie derivative of gauge potentials, whereas it is meaningless to speak about their covariant derivatives
since they are connections. Secondly, we emphasize that $\tilde{\pounds}_{\alpha \xi} \neq \alpha \tilde{\pounds}_{\xi}$, where $\alpha$ is an arbitrary scalar function. Thus, $\tilde{\pounds}_{\xi}$ does not possess the directional property, whence it cannot be interpreted as a covariant derivative. The difference becomes clearer if one takes into account $\tilde{\pounds}_{\xi} T$, for some arbitrary tensor field $T$ which has, at least, one holonomic index. Indeed, such expression depends on the (covariant) derivative of $\xi$, i.e.\ it is only local in $\xi$, whereas $\tilde{\nabla}_\xi$ is always punctual in $\xi$.

\subsubsection*{Algebra properties of the covariant Lie derivatives}

The algebraic properties of the full group of infinitesimal transformations,
and especially of the covariant ones, hinges on the proof of Eq.~\eqref{cov-Lie-d-algebra}. The interplay between the covariant Lie derivative and ordinary gauge variations can be summarized in the commutator
\begin{equation}
 \bigl[\widetilde{\pounds}_\xi,\delta^L_{\alpha}\bigr] = \delta^L_{\xi \cdot D\alpha}\,,
 \qquad 
 \bigl[\widetilde{\pounds}_\xi,\delta^W_{\sigma}\bigr] = \delta^W_{\xi \cdot \partial\sigma}\,,
\end{equation}
An important difference between $\pounds_\xi={\delta}^E_{\xi}$ and $\widetilde{\pounds}_\xi$ is that $\pounds_\xi$ has trivial commutators
\begin{equation}\label{impr-Einstein-no-twist}
 \big[ \delta^E_\xi , \delta^L_{\zeta \cdot \omega} \big] = 0\,, \qquad
 \big[ \delta^E_\xi , \delta^W_{\zeta \cdot S} \big] =0\,,
\end{equation}
instead, the improved transformation does not, as seen in Eq.~\eqref{impr-Einstein-twist} of the main text.

\subsubsection*{Commutator acting on tensors}

First of all, we want to prove the following commutator of two covariant Lie derivatives acting on a Lorentz tensor with Weyl weight $w_A$
\begin{equation}
 \big[\widetilde{\pounds}_\xi, \widetilde{\pounds}_\zeta \big] A^{a\phantom{b}\rho}_{\phantom{a}b\phantom{\rho}\kappa} = \widetilde{\pounds}_{[\xi,\zeta]}A^{a\phantom{b}\rho}_{\phantom{a}b\phantom{\rho}\kappa} + \delta^L_{\mathcal{R}(\xi,\zeta)}A^{a\phantom{b}\rho}_{\phantom{a}b\phantom{\rho}\kappa} + \delta^W_{\mathcal{W}(\xi,\zeta)} A^{a\phantom{b}\rho}_{\phantom{a}b\phantom{\rho}\kappa}\,,
\end{equation}
where the tensor is chosen to have covariant and contravariant, holonomic and anholonomic indices (the extension to an arbitrary number of indices is straightforward).
A further simplification comes from noticing that Lie derivatives, exterior derivatives and contractions are intrinsic operations on a generic manifold, so we can suppress coordinate indices henceforth. Thus, we need to prove that
\begin{equation}
 \big[\widetilde{\pounds}_\xi, \widetilde{\pounds}_\zeta \big] A^a{}_b = \widetilde{\pounds}_{[\xi,\zeta]}A^a{}_b + \delta^L_{\mathcal{R}(\xi,\zeta)} A^a{}_b + \delta^W_{\mathcal{W}(\xi,\zeta)} A^a{}_b \, ,
\end{equation}
where $A^a{}_b$ is a generic $(p,q)$-tensor. We need the following formula, which is easy to show
\begin{equation}\label{res1}
 \pounds_\xi (\zeta \cdot S) - \pounds_\zeta (\xi \cdot S) = dS (\xi,\zeta) + \big[ \xi, \zeta\big] \cdot S\,,
\end{equation}
and is valid for any $1$-form, that is, we can use it for the spin-connection
by replacing $S_\mu$ with $\omega^a{}_{b\mu}$.

Using the definition of covariant Lie derivative, we write
\begin{align}
 \big[\widetilde{\pounds}_\xi, \widetilde{\pounds}_\zeta \big] A^a{}_b & = \left( \pounds_\xi + \delta^L_{\xi\cdot \omega} + \delta^W_{\xi\cdot S} \right) \left(\widetilde{\pounds}_\zeta A^a{}_b \right) - (\zeta \leftrightarrow \xi) \,,
\end{align}
and exploit the gauge and coordinate covariance of $\widetilde{\pounds}_\zeta A^a{}_b$ to obtain
\begin{align}\label{step1}
 \big[\widetilde{\pounds}_\xi, \widetilde{\pounds}_\zeta \big] A^a{}_b & = \pounds_\xi \left(\widetilde{\pounds}_\zeta A^a{}_b\right) + (\xi \cdot \omega)^a{}_c \left(\widetilde{\pounds}_\zeta A^c{}_b\right) \\\nonumber
 & - (\xi \cdot \omega)^c{}_b \left(\widetilde{\pounds}_\zeta A^a{}_c \right) + w_A (\xi \cdot S ) \left(\widetilde{\pounds}_\zeta A^a{}_b \right) - \left( \zeta \leftrightarrow \xi \right)
\end{align}
We focus on the first term on the right hand side and apply the Leibniz rule to the ordinary Lie derivative,
\begin{align}\label{step2}
 \pounds_\xi \left( \widetilde{\pounds}_\zeta A^a{}_b \right) & = \pounds_\xi \left( \pounds_\zeta \right) A^a{}_b + \overline{\pounds_\zeta \left( \pounds_\xi A^a{}_b \right)} + \pounds_\xi \left( \zeta \cdot \omega \right)^a{}_c A^c{}_b + \underline{\left( \zeta \cdot \omega \right)^a{}_c \left( \pounds_\xi A^c{}_b \right)} \\\nonumber
 & - \pounds_\xi \left( \zeta \cdot \omega \right)^c{}_b A^a{}_c - \underline{\left( \zeta \cdot \omega \right)^c{}_b \left( \pounds_\xi A^a{}_c \right)} + w_A \pounds_\xi \left( \zeta \cdot S \right) A^a{}_b + \underline{w_A \left( \zeta \cdot S \right) \left( \pounds_\xi A^a{}_b \right)}\,.
\end{align}
The remaining three terms on the right hand side of Eq.~\eqref{step1} give
\begin{subequations}
 \begin{align}\label{step3.1}
  (\xi \cdot \omega)^a{}_c \left(\widetilde{\pounds}_\zeta A^c{}_b \right) & = (\xi \cdot \omega)^a{}_c \left\{ \underline{\pounds_\zeta A^c{}_b} + \left( \zeta \cdot \omega \right)^c{}_d A^d{}_b - \overline{\left( \zeta \cdot \omega \right)^d{}_b A^c{}_d} + \overline{\overline{w_A \left( \zeta \cdot S \right) A^c{}_b}} \right\} \,, \\ \label{step3.2}
  (\xi \cdot \omega)^c{}_b \left(\widetilde{\pounds}_\zeta A^a{}_c \right) & = (\xi \cdot \omega)^c{}_b \left\{ \underline{\pounds_\zeta A^a{}_c} + \overline{\left( \zeta \cdot \omega \right)^a{}_d A^d{}_c} - \left( \zeta \cdot \omega \right)^d{}_c A^a{}_d + \underline{\underline{w_A \left( \zeta \cdot S \right) A^a{}_c}} \right\} \,, \\ \label{step3.3}
  (\xi \cdot S) \left(\widetilde{\pounds}_\zeta A^a{}_b \right) & = (\xi \cdot S) \left\{ \underline{\pounds_\zeta A^a{}_b} + \overline{\overline{\left( \zeta \cdot \omega \right)^a{}_c A^c{}_b}} - \underline{\underline{\left( \zeta \cdot \omega \right)^c{}_b A^a{}_c}} + w_A \left( \zeta \cdot S \right) A^a{}_b \right\} \,.
 \end{align}
\end{subequations}
Upon antisymmetrization in $\xi \leftrightarrow \zeta$ many terms combine.
All the terms double underlining or overlining cancel out against each others upon antisymmetrization.
Furthermore, the underlined terms in Eqs.~\eqref{step3.1}, \eqref{step3.2} and \eqref{step3.3} with the ordinary Lie derivatives, simplify with the underlined ones in \eqref{step2}. Finally, the last term in \eqref{step3.3} drops, since Weyl symmetry is Abelian.

Thus, using $\left[\pounds_\xi,\pounds_\zeta\right] A^a{}_b = \pounds_{[\xi,\zeta]} A^a{}_b$, we can write
\begin{align}
 \big[\widetilde{\pounds}_\xi, \widetilde{\pounds}_\zeta \big] A^a{}_b  = &\,  \pounds_{[\xi,\zeta]} A^a{}_b + \left[ \pounds_\xi \left( \zeta \cdot \omega \right)^a{}_c - \pounds_\zeta \left( \xi \cdot \omega \right)^a{}_c \right] A^c{}_b\\\nonumber
 & - \left[ \pounds_\xi \left( \zeta \cdot \omega \right)^c{}_b - \pounds_\zeta \left( \xi \cdot \omega \right)^c{}_b \right] A^a{}_c\\\nonumber
 & + w_A \left[ \pounds_\xi \left( \zeta \cdot S \right) - \pounds_\zeta \left( \xi \cdot S \right) \right] A^a{}_b\\\nonumber
 & + \left[ (\xi \cdot \omega)^a{}_c \left( \zeta \cdot \omega \right)^c{}_d - (\zeta \cdot \omega)^a{}_c \left( \xi \cdot \omega \right)^c{}_d \right] A^d{}_b\\\nonumber
 & - \left[ (\xi \cdot \omega)^c{}_b \left( \zeta \cdot \omega \right)^d{}_c - (\zeta \cdot \omega)^c{}_b \left( \xi \cdot \omega \right)^d{}_c \right] A^a{}_d \,.
\end{align}
Further, exploiting the result \eqref{res1}, we get
\begin{align}
 \big[\widetilde{\pounds}_\xi, \widetilde{\pounds}_\zeta \big] A^a{}_b = &\, \pounds_{[\xi,\zeta]} A^a{}_b + \left( \big[ \xi, \zeta \big] \cdot \omega \right)^a{}_c A^c{}_b - \left( \big[ \xi, \zeta \big] \cdot \omega \right)^c{}_b A^a{}_c + \left( \big[ \xi, \zeta \big] \cdot S \right) A^a{}_b \\\nonumber
 & + \, \mathcal{R}(\xi,\zeta)^a{}_c A^c{}_b - \mathcal{R}(\xi,\zeta)^c{}_b A^a{}_c + w_A \mathcal{W} (\xi,\zeta) A^a{}_b \\\nonumber
  = & \, \left( \widetilde{\pounds}_{[\xi, \zeta]} + \delta^L_{\mathcal{R}(\xi,\zeta)} + \delta^W_{\mathcal{W}(\xi,\zeta)} \right) A^a{}_b 
 \, ,
\end{align}
as given in the main text.

\subsubsection*{Commutator acting on connections}

The previous proof holds for Lorentz and Weyl tensor, however, we know that physical fields can transform in a much more general way, e.g., as connections.
It might be unclear whether the same structure of the algebra holds for the connections as well, so we are going to show that the same commutation rule hold for the spin connection. The proof for any other gauge-connection relies on the same steps, even though for a given affine connection $\Gamma^\rho{}_{\nu\mu}$ the same result happens to be a consequence of the tetrad postulate, the Leibniz rule and the fact that $\Gamma^\rho{}_{\nu\mu}$ is Lorentz and Weyl invariant from the onset.

Using the result given in Eq.~\eqref{impr-Einstein} of the main text, we know that
\begin{equation}
 \widetilde{\pounds}_\xi \omega^a{}_{b\mu}= \xi^\nu R^a{}_{b\nu\mu}\,,
\end{equation}
so, we have to prove that
\begin{equation}\label{algebra-spin}
 \big[ \widetilde{\pounds}_\xi, \widetilde{\pounds}_\zeta \big] \omega^a{}_{b\mu} = \big[ \xi,\zeta \big]^\nu R^a{}_{b\nu\mu} - D_\mu (\mathcal{R}^a_{\phantom{a}b}(\xi,\zeta)),
\end{equation}
where $D_\mu$ is the gauge-covariant derivative. Since the Riemann $2$-form and the spin-connection are Weyl invariant, the Weyl variation will not appear in the following equations. We have
\begin{align}\label{algebra-spin-conn1}
 \big[ \widetilde{\pounds}_\xi, \widetilde{\pounds}_\zeta \big] \omega^a{}_{b\mu} = & \, \widetilde{\pounds}_\xi \left( \zeta^\nu R^a{}_{b\nu\mu}  \right) - \widetilde{\pounds}_\zeta \left( \xi^\nu R^a{}_{b\nu\mu}  \right) \\\nonumber
 = &\, \pounds_\xi(\zeta^\nu) R^a{}_{b\nu\mu} + \zeta^\nu \pounds_\xi R^a{}_{b\nu\mu} + \zeta^\nu \delta^L_{\xi\cdot \omega} R^a{}_{b\nu\mu}
 \\\nonumber
 &
 - \, \pounds_\zeta(\xi^\nu) R^a{}_{b\nu\mu} - \xi^\nu \pounds_\zeta R^a{}_{b\nu\mu} - \xi^\nu \delta^L_{\zeta \cdot \omega} R^a{}_{b\nu\mu}\\\nonumber
 = & \, 2 \big[ \xi, \zeta \big]^\nu R^a{}_{b\nu\mu} \, + \\\nonumber
 & + \, \zeta^\nu \xi^\rho \left[ \partial_\rho R^a{}_{b\nu\mu} - \partial_\nu R^a{}_{b\rho\mu} + \omega^a{}_{c\rho} R^c{}_{b\nu\mu} - \omega^c{}_{b\rho} R^a{}_{c\nu\mu} - \omega^a{}_{c\nu} R^c{}_{b\rho\mu} + \omega^c{}_{b\nu} R^a{}_{c\rho\mu} \right]\\\nonumber
 & + \, \zeta^\nu R^a{}_{b\rho\mu} \partial_\nu \xi^\rho + \zeta^\nu R^a{}_{b\nu\rho} \partial_\mu \xi^\rho - \xi^\nu R^a{}_{b\rho\mu} \partial_\nu \zeta^\rho - \xi^\nu R^a{}_{b\nu\rho} \partial_\mu \zeta^\rho
\end{align}
Notice that the first and third terms in the last line can be combined in $[\zeta,\xi]^\nu R^a{}_{b\nu\mu}$, thus canceling the factor $2$ which appears in the commutator of two lines above. Another manipulation comes by adding and subtracting $\zeta^\nu \xi^\rho \partial_\mu R^a{}_{b\rho\nu}$,  $ \zeta^\nu \xi^\rho \omega^a{}_{c\mu} R^c{}_{b\nu\rho}$ and $ - \, \zeta^\nu \xi^\rho \omega^c{}_{b\mu} R^a{}_{c\nu\rho}$. The added terms combine with those in the square brackets to give the second Bianchi identities for the Riemann tensor, $ D_{[\mu} R^a{}_{|b|\nu\rho]} = 0 $.
Instead, the subtracted terms add up with those in the last line of \eqref{algebra-spin-conn1} to give the gauge-covariant derivative, which appears in the right hand side of \eqref{algebra-spin}, thus completing the proof. With hindsight, we notice that the commutation rule applies for all gauge potentials, provided that the second Bianchi identities hold.

\subsubsection*{Jacobi identities}

The improved transformations are a field-dependent generalizations
of a Lie algebra, which is closed if the Jacobi identities hold.
For the algebra to be closed, we thus need to prove that the identities hold. We consider three vector fields $\xi$, $\zeta$ and $\chi$.
We have that
\begin{align}
 {\rm Cycl}_{\xi,\zeta,\chi} \big[ \widetilde{\pounds}_\xi, \widetilde{\pounds}_\zeta \big] \left( \widetilde{\pounds}_\chi A^a{}_b \right) = {\rm Cycl}_{\xi,\zeta,\chi} \widetilde{\pounds}_\xi \left( \widetilde{\pounds}_{[\zeta,\chi]} A^a{}_b + \delta^L_{\mathcal{R}(\zeta,\chi)} A^a{}_b + \delta^W_{\mathcal{W}(\zeta,\chi)} A^a{}_b \right)\,,
\end{align}
where the notation ${\rm Cycl}_{\xi,\zeta,\chi}$ stands for the sum over a cyclic
permutation of the three vectors.
Using that result and exploiting the cyclicity of the sum, we find
\begin{align}
 {\rm Cycl}_{\xi,\zeta,\chi} \big[ \widetilde{\pounds}_\xi , \big[ \widetilde{\pounds}_\zeta , \widetilde{\pounds}_\chi \big] \big] A^a{}_b & = {\rm Cycl}_{\xi,\zeta,\chi} \widetilde{\pounds}_\xi \left( \widetilde{\pounds}_{[\zeta,\chi]} A^a{}_b + \delta^L_{\mathcal{R}(\zeta,\chi)} A^a{}_b + \delta^W_{\mathcal{W}(\zeta,\chi)} A^a{}_b \right)\\\nonumber
 & - {\rm Cycl}_{\xi,\zeta,\chi} \big[ \widetilde{\pounds}_\zeta, \widetilde{\pounds}_\chi \big] \left( \widetilde{\pounds}_\xi A^a{}_b \right) = 0\,,
\end{align} 
that proves the Jacobi identities for the covariant Lie derivatives.



\bibliographystyle{chetref}
\bibliography{biblio}

\end{document}